\documentclass[onecolumn,draftcls]{IEEEtran}

\usepackage{latexsym}
\usepackage{amssymb}
\usepackage{bm}
\usepackage[dvips]{graphics}
\usepackage{graphicx}
\usepackage{psfrag}
\usepackage{amsfonts,amsmath,amssymb,hyperref}
\usepackage{algorithm,algorithmic}

\usepackage{dsfont,color,subfigure}

\usepackage{xspace}
\usepackage{bbm}

\DeclareMathOperator*{\argmin}{arg\,min}

\newcommand{\Ac}{\mathcal{A}}

\newcommand{\Cc}{\mathcal{C}}

\newcommand{\Ec}{\mathcal{E}}

\newcommand{\Hc}{\mathcal{H}}

\newcommand{\Sc}{\mathcal{S}}

\newcommand{\Xc}{\mathcal{X}}
\newcommand{\Yc}{\mathcal{Y}}

\def\a{\alpha}
\def\b{\beta}

\def\e{\epsilon}

\DeclareMathOperator\E{E}
\let\P\relax
\DeclareMathOperator\P{P}

\newcommand{\Bern}{\mathrm{Bern}}

\def\textiid{i.i.d.\@\xspace}
\newcommand\iid{\ifmmode\text{ i.i.d. } \else \textiid \fi}

\newcommand{\mb}{\mathbf{m}}
\newcommand{\ub}{\mathbf{u}}

\newcommand{\bb}{\mathbf{b}}

\newcommand{\pb}{\mathbf{p}}

\begin{document}

\title{Rate-Distortion via Markov Chain Monte Carlo}

\author{\authorblockN{Shirin Jalali\authorrefmark{1} and Tsachy Weissman\authorrefmark{1}\authorrefmark{2},}
\authorblockA{\authorrefmark{1}Department of Electrical
Engineering, Stanford University, Stanford, CA 94305, \{shjalali, tsachy\}@stanford.edu}
} \maketitle

\newtheorem{property}{Property}
\newtheorem{question}{Question}
\newtheorem{claim}{Claim}
\newtheorem{guess}{Conjecture}
\newtheorem{definition}{Definition}
\newtheorem{fact}{Fact}
\newtheorem{assumption}{Assumption}
\newtheorem{theorem}{Theorem}
\newtheorem{lemma}{Lemma}
\newtheorem{ctheorem}{Corrected Theorem}
\newtheorem{corollary}{Corollary}
\newtheorem{proposition}{Proposition}
\newtheorem{example}{Example}
\newtheorem{remark}{Remark}

\renewcommand{\algorithmicrequire}{\textbf{Input:}}

\renewcommand{\algorithmicensure}{\textbf{Output:}}

\begin{abstract}

We propose an approach to lossy source coding, utilizing ideas from Gibbs sampling, simulated annealing, and Markov Chain Monte Carlo (MCMC). The idea is to sample a reconstruction sequence from a Boltzmann distribution associated with an energy function that incorporates the distortion between the source and reconstruction, the compressibility of the reconstruction, and the point sought on the rate-distortion curve. To sample from this distribution, we use a `heat bath algorithm': Starting from an initial candidate reconstruction (say the original source sequence), at every iteration, an index $i$ is chosen and the $i^{\rm th}$ sequence component is replaced by drawing from the conditional probability distribution for that component given all the rest. At the end of this process, the encoder conveys the reconstruction to the decoder using universal lossless compression.

The complexity of each iteration is independent of the sequence length and only linearly dependent on a certain context parameter (which grows sub-logarithmically with the sequence length). We show that the proposed algorithms achieve optimum rate-distortion performance in the limits of large number of iterations, and sequence length, when employed on any stationary ergodic source. Experimentation shows promising initial results.

Employing our lossy compressors on noisy data, with appropriately chosen distortion measure and level, followed by a simple de-randomization operation, results in a family of denoisers that compares favorably (both theoretically and in practice) with other MCMC-based schemes, and with the Discrete Universal Denoiser (DUDE).

\end{abstract}


\section{introduction}\label{sec: intro}

Consider the basic setup of lossy coding of a stationary ergodic
source $\mathbf{X}=\{X_i: i \geq 1\}$. Each source output block of length
$n$, $X^n$, is mapped to an index $f_n(X^n)$ of $nR$ bits, where $R$ can be either constant (fixed-rate coding) or depend on the block that is coded (variable-rate coding). The index $f_n(X^n)$ is then losslessly transmitted to the decoder, and is decoded to a
reconstruction block $\hat{X}^n = g_n(f_n(X^n))$. Two main performance measures for a lossy coding scheme $\mathcal{C}=(f_n,g_n,n)$ are the following: i) distortion $D$ defined as average expected distortion between source and reconstruction blocks, i.e., 
\begin{equation}
D\triangleq\E d_n(X^n,\hat{X}^n)\triangleq\frac{1}{n}\sum\limits_{i=1}^n \E d(X_i,\hat{X}_i),
\end{equation}
where $d:\mathcal{X}\times\mathcal{X}\rightarrow {\mathds{R}}^+$ is a single-letter distortion measure, and ii) rate $R$ defined as the average expected number of bits per source symbol, i.e., $\E[R]$.  For any $D\geq0$, and stationary process $\mathbf{X}$ the minimum achievable rate (cf.~\cite{cover} for exact definition of achievability)  is characterized as  \cite{Shannon48}, \cite{Gallager}, \cite{book:Berger}
\begin{equation} \label{eq: rate-distortion function}
R(D,\mathbf{X})=\lim\limits_{n\rightarrow\infty}\min\limits_{p(\hat{X}^n|X^n):\E d_n(X^n,\hat{X}^n)\leq D}\frac{1}{n}I(X^n;\hat{X}^n).
\end{equation}
For the case of lossless compression, we know that the minimum required rate is the entropy rate of the source,
i.e.~$\bar{H}(\mathbf{X})\triangleq\lim\limits_{k\rightarrow\infty} H(X_0|X_{-k}^{-1})$, and there are known implementable \emph{universal} schemes, such as Lempel-Ziv coding \cite{LZ} and arithmetic coding
\cite{arith_coding}, that are  able to describe any stationary
ergodic source at rates as close as desired to the entropy rate of
the source without any error. In contrast to the situation of lossless compression, neither the explicit solution of (\ref{eq: rate-distortion function}) is known for a general source
(not even for a first-order Markov source \cite{ISIT_JW}), nor are there known
practical schemes that universally achieve  the rate-distortion
curve.

One possible intuitive explanation for this sharp dichotomy is as follows. The essence of universal lossless compression algorithms is \emph{learning} the source distribution, and the difference between various coding algorithms is in different efficient methods through which they  accomplish this goal. Universal lossy compression, on the other hand, intrinsically consists of two components: quantization and lossless compression. This breakdown can be explained more clearly by the following characterization of the rate-distortion function \cite{GrayN:75}:
\begin{align}
R(D,\mathbf{X})=\inf \{\bar{H}(\mathbf{Z}): \; \E d(X_1,Z_1)\leq D\},\label{eq: r-d alternative rep}
\end{align}
where the infimum in over jointly stationary ergodic processes with $\mathbf{X}$. This alternative representation suggests that for coding a process $\mathbf{X}$ one should quantize it, either implicitly or explicitly, to another process $\mathbf{Z}$, which is sufficiently close to it but more compressible, and then compress process $\mathbf{Z}$ via a universal lossless compression algorithm. The quantization step in fact involves a search over the space of all jointly stationary ergodic processes, and explains to some extent the reason why universal lossy compression is more intricate than universal lossless compression.

In this paper, we present a new approach to implementable lossy  source coding,
which borrows two well-known tools from statistical physics and computer science, namely Markov Chain
Monte Carlo (MCMC) methods, and simulated annealing  \cite{simulated_annealing_1,simulated_annealing_2}. MCMC methods refer to a
class of algorithms that are designed to generate samples of a given
distribution through generating a Markov chain having the desired
distribution as its stationary distribution. MCMC methods include a
large number of algorithms; For our application, we use  Gibbs
sampler \cite{Gibbs_sampler} also known as the {\it heat bath}
algorithm,  which is well-suited  to the case where the desired
distribution is hard to compute, but the conditional distributions
of each variable given the rest are easy to work out.

The second required tool is simulated annealing which is a
well-known method in discrete optimization problems. Its goal is to find the
the minimizing state $s_{\textmd{min}}$ of a function $f(s)$ over a set of possibly huge number of states $\mathcal{S}$, i.e., $s_{\textmd{min}}= \argmin\limits_{s\in\mathcal{S}} f(s)$. In order to do simulated annealing, a
sequence of probability distributions $p_1,p_2,\ldots$ corresponding
to the temperatures $T_1>T_2>\ldots$, where $T_i\rightarrow 0$ as
$i\rightarrow\infty$, and a sequence of positive integers $N_1$,
$N_2$, $\ldots$, are considered. For the first $N_1$ steps, the
algorithm runs one of the relevant MCMC methods in an attempt to
sample from distribution $p_1$. Then, for the next $N_2$ steps, the
algorithm, using the output of the previous part as the initial
point, aims to sample from $p_2$, and so on. The probability
distributions are designed such that: 1) their output, with high
probability, is the minimizing state $s_{\textmd{min}}$, or one of
the states close to it, 2) the probability of getting the minimizing
state increases as the temperature drops. The probability
distribution that satisfies these characteristics, and is almost
always used, is the Boltzman distribution $p_{\beta}(s)\propto
e^{-\beta f(s)}$, where $\beta \propto \frac{1}{T}$. It can be
proved that using Boltzman distribution, if the temperature drops
slowly enough, the probability of ultimately getting the minimizing
state as the output of the algorithm approaches one \cite{Gibbs_sampler}. Simulated annealing has been suggested before in the
context of lossy compression, either as a way for approximating the
rate distortion function (i.e., the optimization problem involving
minimization of the mutual information) or as a  method for
designing the codebook in vector quantization \cite{deterministic_annealing,SA_codebook_design}, as an
alternative to the conventional generalized Lloyd algorithm (GLA)
\cite{GLA}.
In contrast, in this paper  we use the simulated annealing approach
to obtain a particular reconstruction sequence, rather than a whole
codebook.

Let us briefly describe how the new algorithm codes a source
sequence $x^n$. First, to each reconstruction block $y^n$, it
assigns an \emph{energy}, $\Ec(y^n)$, which is a linear combination of
its conditional empirical entropy, to be defined formally in the
next section, and its distance from the source sequence $x^n$. Then,
it assumes a Boltzman probability distribution over the
reconstruction blocks as $p(y^n) \propto e^{-\beta \Ec(y^n)}$, for
some $\beta>0$, and tries to generate $\hat{x}^n$  from this
distribution using Gibbs sampling \cite{Gibbs_sampler}. As we will
show, for $\beta$ large enough, with high probability the
reconstruction block of our algorithm would satisfy
$\Ec(\hat{x}^n)\approx \min \Ec(y^n)$. The encoder will output
${\sf\footnotesize  LZ}(\hat{x}^n)$, which is the Lempel-Ziv \cite{LZ} description of
$\hat{x}^n$. The decoder, upon receiving ${\sf\footnotesize  LZ}(\hat{x}^n)$,
reconstructs $\hat{x}^n$ perfectly.

In this paper, instead of working at a fixed rate or at a fixed
distortion, we are fixing the slope. A fixed slope rate-distortion
scheme, for a fixed slope $s=-\alpha<0$, looks for the coding scheme that
minimizes $R+\alpha D$, where as usual $R$ and $D$ denote the rate
and the average expected  distortion respectively. In comparison to
a given coding scheme of rate $R$ and expected distortion $D$, for
any $0<\delta<R-R(D,\mathbf{X})$, there exists a code which works at
rate $R(D,\mathbf{X})+\delta$ and has the same average expected
distortion, and consequently a lower cost. Therefore, it follows
that any point that is optimal in the fixed-slope setup corresponds
to a point on the rate-distortion curve.

\subsection{Prior work}
The literature on universal lossy compression can be divided into two main categories: existence proofs and algorithm designs. The early works in this area were more about proving the existence of a family of codes $(n,f_n,g_n)$ that achieves the optimal performance, $R(D,\mathbf{X})$, asymptotically for any stationary ergodic process \cite{Sakrison:70,Ziv:72,NeuhoffG:75,NeuhoffS:78,Ziv:80,GarciaN:82}. After the existence of the so-called \emph{universal} codes were shown, the next step was finding such algorithms. We will here briefly review some of the work on the latter. This section is not meant to be a thorough review of the literature on universal lossy compression algorithms, but just a brief overview of some of the more famous results to the knowledge of the authors.

One popular trend in finding universal lossy compression algorithms has been extending universal lossless compression algorithms to the lossy case. As an example of such attempts is the work by Cheung and Wei \cite{CheungS:90} who extended the move-to-front transform \cite{BentleyS:86}. There has also been a lot of attempt on extending the string-matching ideas used in the well-known Lempel-Ziv coding to the lossy case: Morita and Kobayashi \cite{MoritaK:89} proposed a lossy version of LZW algorithm and Steinberg and Gutman \cite{SteinbergG:93} suggested a fixed-database lossy compression algorithms based on string-matching. These algorithms have the same spirit of LZ coding, and similar to the LZ code are easy to implement. However, all these extensions, as were later shown by Yang and Kieffer  \cite{YangK:98}, are suboptimal even for memoryless sources. Another suboptimal but practical universal lossy compression algorithm based on approximate pattern matching is the work of Luczak and Szpankowski \cite{LuczakS:97}.

Zhang and Wei \cite{ZhangW:96} proposed an online universal lossy data compression algorithm, called `gold-washing', which involves continuous codebook refinement. The algorithm is called online meaning that the codebook in constructed simultaneously  by the encoder and the decoder as the source symbols arrive, and no codebook is shared between the two before the coding starts. Most of the previously mentioned algorithms fall into the class of online algorithms as well.

More recently, a new lossy version of LZ algorithm has been proposed by Kontoyiannis \cite{kontoyiannis_1} which instead of using a fixed database which has the same distribution as the source, employs multiple databases. The encoder is allowed to choose one of the databases at each step. These multiple databases essentially let the encoder tune the reconstruction distribution gradually to the optimal distribution that corresponds to the source distribution. It is a fixed-distortion code and is shown to be optimal, at least, for memoryless sources.

There are also universal lossy compression algorithms that are interesting from a theoretical point-of-view, but infeasible to be implemented because of their huge computational complexity. One can refer to the works by Ornstein and Shields \cite{OrnsteinS:90}, Yang and Kieffer \cite{YangK:96}, and more recently Neuhuff and Shields \cite{NeuhoffS:98} for examples of such results.

As mentioned earlier in this paper the encoder, instead of fixing rate or distortion,  fixes the slope. The idea of fixed-slope universal lossy compression was first proposed by Yang, Zhang and Berger  in \cite{YangZ:97}. In their paper, they first  propose an exhaustive search coding algorithm which is very similar to the algorithm proposed propose in Section \ref{exhaustive}. After establishing its universality for lossy compression of stationary ergodic sources, they suggest some heuristic  approach  for finding an approximation to its solution. In our case, the special structure of our cost function enables us to employ simulated annealing plus Gibbs sampling to approximate its minimizer. 

For the non-universal setting, specifically the case of lossy compression of an i.i.d.~source with a known distribution, there is an ongoing progress towards designing codes that get very close to the optimal performance  \cite{non-universal_R_D_coder_ref1,non-universal_R_D_coder_ref2,R_D_coder_ref3,R_D_coder_ref4}.

\subsection{Paper organization}

The organization of the paper is as follows.  In Section \ref{sec:notation}, we set up the notation. Section \ref{exhaustive} describes an exhaustive search scheme for fixed-slope lossy compression which universally achieves the rate-distortion curve for any stationary ergodic source.  Section \ref{sec: universal lossy} describes our new universal MCMC-based lossy coder, and Section \ref{sec: sliding} presents another version of the algorithm for finding sliding-block codes which again universally attain the rate-distortion bound. Section \ref{sec: simulations} gives some simulations results. Section \ref{sec: denoising} describes the application of the algortihm introduced in Section \ref{sec: universal lossy} to universal compression-based denoising. Finally, Section \ref{conclusion} concludes the paper with a discussion of some future directions.

\section{Notation}\label{sec:notation}

Let $\mathbf{X}=\{X_i;\forall\; i\in\mathds{N}^{+}\}$ be a
stochastic process defined on a probability space
$(\mathbf{X},\Sigma,\mu)$, where  $\Sigma$ denotes the
$\sigma$-algebra generated by cylinder sets $\Cc$, and $\mu$ is
a probability measure defined on it. For a process
$\mathbf{X}$, let $\Xc$ denote the alphabet of $X_i$, which is
assumed to be finite. The shift operator
$T:\Xc^{\infty}\to\Xc^{\infty}$ is defined by
\[
(T\mathbf{x})_n=x_{n+1},\quad\mathbf{x}\in\Xc^{\infty},n\geq1.
\]
For a stationary process $\mathbf{X}$, let
$\bar{H}(\mathbf{X})$ denote its entropy rate defined as
$\bar{H}(\mathbf{X})=\lim\limits_{n\to\infty}H(X_{n+1}|X^n)$.   

Calligraphic letters, $\Xc$, $\Yc$, etc, are always assumed to refer to sets, and usually represent the alphabet sets of random variables. The size of a set $\Ac$ is denoted by $|\Ac|$. Specifically, let $\Xc$ and $\hat{\Xc}$ denote the source and reconstruction alphabets respectively.

For $y^n\in\Yc^n$, define the matrix
$\mathbf{m}(y^n)\in\mathds{R}^{|\Yc|}\times\mathds{R}^{|\Yc|^{k}}$
to be $(k+1)^{\rm th}$ order empirical count of $y^n$, i.e.,
its $(\b, \bb)^{\rm th}$ element is  defined as
\begin{equation}\label{eq: empirical count matrix}
m_{\b,\bb}(y^n) = \frac{1}{n-k} \left| \left\{ k+1 \leq i \leq n :
y_{i-k}^{i-1} = \bb, y_i=\b]    \right\}\right|,
\end{equation}
where $\bb\in\Yc^k$, and $\b\in\Yc$.
Let $H_k(y^n)$ denote the conditional empirical entropy of order $k$ induced by $y^n$, i.e.,
\begin{equation}\label{eq: emp cond distribution}
   H_k (y^n) = H(Y_{k+1} | Y^{k}) ,
\end{equation}
where $Y^{k+1}$ on the right hand side of (\ref{eq: emp cond
distribution}) is distributed according to
\begin{equation}\label{eq: empirical distribution}
   \P (Y^{k+1} = [\bb,\b]) = m_{\b,\bb}(y^n).
\end{equation}
For a vector
$\mathbf{v}= (v_1, \ldots , v_\ell)^T$ with non-negative
components, we let $\mathcal{H}(\mathbf{v})$ denote the entropy
of the random variable whose probability mass function (pmf) is
proportional to $\mathbf{v}$. Formally,
\begin{equation}\label{eq: single letter ent functional defined}
\mathcal{H} (\mathbf{v}) = \left\{ \begin{array}{cc}
                           \sum\limits_{i=1}^\ell \frac{v_i}{\| \mathbf{v}
\|_1}  \log \frac{\| \mathbf{v} \|_1}{v_i} &  \mbox{ if }  \mathbf{v}
\neq (0, \ldots , 0)^T \\
                           0 & \mbox{ if } \mathbf{v}  = (0, \ldots , 0)^T,
                         \end{array}
\right.
\end{equation}
where $0\log(0) = 0$ by convention.
The conditional empirical entropy in \eqref{eq: emp cond
distribution} can be expressed as a function of $\mb(y^n)$ as
follows
\begin{equation}\label{eq: alternative representation of Hk}
H_k (y^n) = \frac{1}{n} \sum_{\bb} \mathcal{H} \left(
\mb_{\cdot,\bb}(y^n) \right) \mathbf{1}^T \mb_{\cdot,\bb}(y^n),
\end{equation}
where $\mathbf{1}$ and  $\mb_{\cdot,\bb}(y^n)$ denote the
all-ones column vector of length $|\Yc|$, and the column in
$\mb(y^n)$ corresponding to $\bb$ respectively.

For vectors $\mathbf{u}$ and $\mathbf{v}$ both is $\mathds{R}^n$, let $\|\ub-\mathbf{v}\|_1$ denote the $\ell_1$ distance between $\ub$ and $\mathbf{v}$, defined as follows
\begin{align}
\|\ub-\mathbf{v}\|_1 = \sum\limits_{i=1}^n|u_i-v_i|.
\end{align}
Also the total variation between the two vectors is defined as
\begin{align}
\|\ub-\mathbf{v}\|_{\rm TV}=\frac{1}{2}\|\ub-\mathbf{v}\|_1.
\end{align}

\section{An exhaustive search scheme for fixed-slope compression}\label{exhaustive}

Consider the following scheme for lossy source coding at a fixed slope
$\a > 0$. For each source sequence $x^n$ let the reconstruction
block $\hat{x}^n$ be
\begin{equation}\label{eq: reconstruction via exhaustion for fixed slope s}
    \hat{x}^n = \mbox{arg} \min_{y^n} \left[  H_k (y^n) + \a  d_n (x^n,
    y^n)  \right].
\end{equation}
The encoder, after computing $\hat{x}^n$, losslessly conveys it to
the decoder using {\sf\footnotesize  LZ} compression.
\begin{theorem} \label{th: the exhaustive search scheme}
Let $\mathbf{X}$ be a stationary ergodic source, let
$R(D,\mathbf{X})$ denote its rate distortion function,  and let
$\hat{X}^n$ denote the reconstruction using the above scheme on
$X^n$. Then
\begin{equation}\label{eq: achieving optimal point on the rd curve} \frac{1}{n} \ell_{{\sf\footnotesize  LZ}} (\hat{X}^n) + \a d_n (X^n,   \hat{X}^n )  \stackrel{n \rightarrow \infty}{\longrightarrow}  \min_{D \geq 0} \left[ R(D,\mathbf{X}) + \a D \right],\;\;{\rm a.s.}
\end{equation}
\end{theorem}
In words, the above scheme universally attains the optimum rate-distortion performance at slope $\a$ for any stationary ergodic process. The drawback of the described algorithm is its computational complexity; It involves exhaustive search among the set of all possible reconstructions. The size of this set is $|\hat{\Xc}|^n$ which grows exponentially fast with $n$.\\
\begin{remark} \label{remark:1}Although the exhaustive search algorithm described above is very similar to the generic algorithm proposed in \cite{YangZ:97}, they are in fact different. The algorithm proposed in \cite{YangZ:97} is as follows
\begin{equation}\label{eq: YangZhangAlg}
    \hat{x}^n = \mbox{arg} \min_{y^n} \left[ \frac{1}{n}l(y^n) + \a  d_n (x^n, y^n)  \right],
\end{equation}
where $l(y^n)$ is the length of the binary codeword assigned to $y^n$ by some universal lossless compression algorithm. From this definition, $l(y^n)$ should satisfy the following two conditions:
\begin{enumerate}
\item For any $n\in\mathds{N}$,
\[
\sum\limits_{y^n\in\Yc^n}2^{-l(y^n)}\leq 1.
\]
\item For any stationary ergodic process $\mathbf{X}$,
\begin{align}
\lim\limits_{n\to\infty}\frac{1}{n}l(X^n)=\bar{H}(\mathbf{X}),\;\;{\rm a.s.}
\end{align}
\end{enumerate}
But conditional empirical entropy, $H_k(\cdot)$,  is not a length function ($\sum\limits_{y^n}2^{-nH_k(y^n)}\geq 2^{-nH_k(0,\ldots,0)} + 2^{-nH_k(1,\ldots,1)}=2$, for any $k$ and $n$). Hence, the algorithm proposed above is not an special case of the generic algorithm proposed in \cite{YangZ:97}.
\end{remark}

\begin{remark}  \label{remark:2} Although as described in Remark \ref{remark:1}, $H_k(\cdot)$ is not a length function itself,  it has a close connection to length functions, specifically to $\ell_{\rm LZ}(\cdot)$. This link is described bt Ziv inequality \cite{Ziv_inequality} which states that if $k_n = o( \log n)$, then for any  $\epsilon>0$, there exists $N_{\epsilon}\in\mathds{N}$ such that for \emph{any} individual infinite-length sequence $\mathbf{y}=(y_1,y_2,\ldots)$ and any $n\geq N_{\epsilon}$,
\begin{equation} \label{eq: Ziv ineq remark}
 \left[ \frac{1}{n} \ell_{{\sf\footnotesize  LZ}} (y^n) -H_{k_n}(y^n) \right] \leq
 \epsilon.
\end{equation}
As described in Section \ref{sec: intro}, the process of universal lossy compression can be divided into two steps: quantization and universal lossless compression. The second step which involves universal lossless compression of the quantized sequence is extensively studied in the literature already and can be done efficiently using existing coders. Hence in this paper we focus on the first step, and try to show that it can be done efficiently via simulated annealing. 
\end{remark}

\begin{proof}[Proof of Theorem \ref{th: the exhaustive search scheme}]
From part (1) of Theorem 5 in \cite{YangZ:97},
\begin{align}
\liminf\limits_{n\to\infty} \left[ \frac{1}{n} \ell_{{\sf\footnotesize  LZ}} (\hat{X}^n) + \a  d (X^n, \hat{X}^n ) \right] \geq \min_{D \geq 0} \left[ R(D,\mathbf{X}) +\a  D \right]\;\;{\rm a.s.}\label{eq: lower bound on cost function}
\end{align}
which says that the probability that a sequence of codes asymptotically beats  the fundamental rate-distortion limit is zero.

In order to establish the upper bound, we split the cost function into two terms as follows
\begin{align}
\left[\frac{1}{n} \ell_{{\sf\footnotesize  LZ}} (\hat{X}^n) + \a  d
(X^n,\hat{X}^n) \right]&= \left[\frac{1}{n} \ell_{{\sf\footnotesize  LZ}} (\hat{X}^n) -H_{k_n}(\hat{X}^n)+H_{k_n}(\hat{X}^n) + \a d (X^n,\hat{X}^n ) \right],\\
&=  \left[\frac{1}{n} \ell_{{\sf\footnotesize  LZ}} (\hat{X}^n)-H_{k_n}(\hat{X}^n)\right] +
 \left[H_{k_n}(\hat{X}^n) + \a  d (X^n,\hat{X}^n )
\right].\label{eq: thm 1: split cost}
\end{align}
From \cite{Ziv_inequality}, for $k_n = o( \log n)$ and any given $\epsilon>0$, there exists
$N_{\epsilon}\in\mathds{N}$ such that for \emph{any} individual infinite-length sequence $\mathbf{\hat{x}}=(\hat{x}_1,\hat{x}_2,\ldots)$
and any $n\geq N_{\epsilon}$,
\begin{equation} \label{eq: Ziv ineq}
 \left[ \frac{1}{n} \ell_{{\sf\footnotesize  LZ}} (\hat{x}^n) -H_{k_n}(\hat{x}^n) \right] \leq
 \epsilon.
\end{equation}

Consider an arbitrary point $(R(D,\mathbf{X}),D)$ on the rate-distortion curve corresponding to source $\mathbf{X}$. Then for any $\delta>0$ there exists a process $\tilde{\mathbf{X}}$ such that $(\mathbf{X},\mathbf{\tilde{X}})$ are jointly stationary  ergodic, and moreover \cite{GrayN:75}
\begin{enumerate}
\item $\bar{H}(\tilde{\mathbf{X}})\leq R(D,\mathbf{X}),$
\item $\E d(X_0,\tilde{X}_0)\leq D+\delta$.
\end{enumerate}
Now since for each source block $X^n$, the reconstruction block $\hat{X}^n$ is chosen to minimize $H_k(\hat{X}^n)+\a d(X^n,\hat{X}^n)$, we have
\begin{align}\label{eq: compare block and SB}
H_{k_n}(\hat{X}^n)+\a d(X^n,\hat{X}^n) &\leq H_{k_n}(\tilde{X}^n)+\a d(X^n,\tilde{X}^n).
\end{align}
For a fixed $k$, from the definition of the $k^{th}$ order entropy, we have
\begin{equation}
H_k(\tilde{X}^n) = \frac{1}{n} \sum\limits_{\ub\in\hat{\Xc}^k} \mathbf{1}^T \mb_{\cdot,\ub}(\tilde{X}^n) \mathcal{H} \left( \mathbf{m}_{\cdot,\ub}(\tilde{X}^n) \right),\label{eq1}
\end{equation}
where
\begin{align}
\frac{1}{n}\mathbf{m}_{u_{k+1},\ub} (\tilde{X}^n) &= \frac{1}{n}\sum_{i=1}^n \mathds{1}_{\tilde{X}_{i-k}^i=u^{k+1}} \\
&\stackrel{n \rightarrow \infty}{\longrightarrow} \P\left( \tilde{X}^0_{-k}=u^{k+1} \right),\quad \textmd{w.p.1.} \label{eq: convergance of count vector}
\end{align}
Therefore, combining (\ref{eq1}) and (\ref{eq: convergance of count vector}), as $n$ goes to infinity, $H_k(\tilde{X}^n)$ converges to $H(\tilde{X}_0|\tilde{X}^{-1}_{-k})$ with probability one. It follows from the monotonicity of $H_k(\hat{x}^n)$ in $k$, (\ref{eq: compare block and SB}), and the convergence we just established that for any $\hat{x}^n$ and any $k$,
\begin{align}\label{eq: bounding cost with condistional entropy}
H_{k_n}(\hat{X}^n)+\a d(X^n,\hat{X}^n)\leq H(\tilde{X}_0|\tilde{X}_{-k}^{-1})+\epsilon+\a d(X^n,\tilde{X}^n), \quad\textmd{eventually a.s.}
\end{align}
On the other hand
\begin{align}
d(\tilde{X}^n,X^n)&=\frac{1}{n}\sum_{i=1}^{n}d(X_i,\tilde{X}_i)\stackrel{n\rightarrow \infty}{\longrightarrow} \E d(\tilde{X_0},X_0) \leq D+\delta.
\end{align}
Combining (\ref{eq: Ziv ineq}) and (\ref{eq: bounding cost with condistional entropy}) yields
\begin{align}
\limsup\limits_{n\rightarrow \infty}\left[\frac{1}{n}\ell_{{\sf\footnotesize  LZ}}(\hat{X}^n)+\a d(X^n,\hat{X}^n)\right]\leq H(\tilde{X}_0|\tilde{X}_{-k}^{-1})+2\epsilon + \a (D+\delta)\;\;{\rm a.s.}
\end{align}
The arbitrariness of $k$, $\epsilon$ and $\delta$ implies
\begin{align}
\limsup\limits_{n\rightarrow\infty} \left[\frac{1}{n}\ell_{{\sf\footnotesize  LZ}}(\hat{X}^n)+\a d(X^n,\hat{X}^n)\right] \leq  R(D,\mathbf{X})+\a D,\;\;{\rm a.s.}
\end{align}
for any $D\geq0$. Since the point $(R(D,\mathbf{X}),D)$ was also chosen arbitrarily, it follows that
\begin{align}
\limsup\limits_{n\rightarrow\infty}  \left[\frac{1}{n}\ell_{{\sf\footnotesize  LZ}}(\hat{X}^n)+\a d(X^n,\hat{X}^n)\right] \leq \min\limits_{D\geq0} [R(D,\mathbf{X})+\a D], \label{eq: upper bound on cost function}
\end{align}
Finally, combining (\ref{eq: lower bound on cost function}), and (\ref{eq: upper bound on cost function}) we get the desired result:
\begin{align}
\lim\limits_{n\rightarrow\infty} \left[\frac{1}{n}\ell_{{\sf\footnotesize  LZ}}(\hat{X}^n)+\a d(X^n,\hat{X}^n)\right] = \min\limits_{D\geq0} [R(D,\mathbf{X})+\a D].
\end{align}

\end{proof}

\begin{remark}
As mentioned above, since $H_k(\cdot)$ is not itself a length function, there is a difference between the algorithm mentioned here, and the one proposed in  \cite{YangZ:97}. However, as we will argue shortly, one can establish a connection between the two, and derive the following result directly from the theorem proved in \cite{YangZ:97}:
\begin{equation}\label{eq: thm1_new_form}
H_k(\hat{X}^n) + \a d_n (X^n,   \hat{X}^n )  \stackrel{n \rightarrow \infty}{\longrightarrow}  \min_{D \geq 0} \left[ R(D,\mathbf{X}) + \a D \right],\;\;{\rm a.s.},
\end{equation}
where again $\hat{X}^n$ is a minimizer of \eqref{eq: reconstruction via exhaustion for fixed slope s}. 

Consider the following entropy coding scheme for describing a sequence $y^n\in\Yc^n$. First, divide $y^n$ into $|\Yc|^k$ subsequences $\{y^{n_{\bb}}_{\bb}\}_{\bb\in\Yc^k}$. Each subsequence corresponds to a vector $\bb\in{\Yc}^k$, and consists of those symbols in $y^n$ which are preceded by $\bb$. From our definitions,
\[
n_{\bb}=n\sum\limits_{\b\in\Yc}m_{\b,\bb}.
\]
Now describing the sequence $y^n$ can be done by describing the mentioned subsequences to the decoder separately. Note that the decoder can merge  the subsequences  and form the original sequence $y^n$ easily if it knows  the first $k$ symbols as well. For describing the subsequences, we first send the matrix $\mb(y^n)$ to the decoder. For doing this at most $|\Yc|^{k+1}\lceil \log n \rceil$ bits are required. After having access to the matrix $\mb$, for each subsequence, the decoder finds its length $n_{\bb}$ and also  the number of occurrences of each symbol within it. Then, since there only exists  
\begin{align}
\left(  \begin{array}{lcr}
   & n_{\bb} & \\
  nm_{\a_1,\bb}, & \ldots &, n m_{\a_N,\bb}
   \end{array}\right),
\end{align}
such sequences,  the encoder is able to describe the sequence of interest within this set  by just sending its  index. But from Stirling approximation, i.e.,
\[
n!=\sqrt{2\pi n}\left(\frac{n}{e}\right)^n e^{\lambda_n},
\]
where $\frac{1}{12n+1} \leq \lambda_n \leq \frac{1}{12n}$, it follows that the required number of bits for sending the index can be written as
\begin{align}
\log \left(\begin{array}{lcr}
   & n_{\bb} & \\
  nm_{\a_1,\bb}, & \ldots &, n m_{\a_N,\bb}
   \end{array}\right) = n\mathcal{H}(\mb_{\cdot,\bb})+n\eta(k,n),
\end{align}
where $\eta(k,n)=o(1)$ and does not depend on $\bb$ or $y_{\bb}^{n_{\bb}}$. Denoting the overall number of bits required by this coding scheme for coding the sequence $y^n$ by $l_e(y^n)$ it follows that the number of bits per symbol is  
\begin{align}
\frac{1}{n}l_e(y^n) & = \sum\limits_{\bb}\frac{1}{n}\mathcal{H}(\mb_{\cdot,\bb}) + \frac{{|\Yc|}^k\eta(k,n)+|\Yc|^{k+1}(\log n+1)}{n} \nonumber\\
& = H_k(y^n)+\zeta(k,n),
\end{align}
where $\zeta(k,n)\triangleq\frac{|\mathcal{\Yc}|^k\eta(k,n)+|\Yc|^{k+1}(\log n+1)}{n}=o(n)$ which again does not depend on $y^n$. 
From our construction, clearly $l_e(\cdot)$ is a length function. Moreover, since $\zeta(k,n)$ does not depend on $y^n$,
\begin{align}
\argmin\limits_{y^n}[\frac{1}{n}l_e(y^n)+\alpha d_n(x^n,y^n)]&= \argmin\limits_{y^n}[H_k(y^n)+\zeta(k,n)+\alpha d_n(x^n,y^n)],\nonumber\\
&=\argmin\limits_{y^n}[H_k(y^n)+\alpha d_n(x^n,y^n)].
\end{align}
Therefore,
\begin{align}
\lim\limits_{n\to\infty}[\frac{1}{n}l_e(y^n) + \a d_n (X^n,   \hat{X}^n )] & = \lim\limits_{n\to\infty} [H_k(\hat{X}^n) + \a d_n (X^n,   \hat{X}^n )] \nonumber \\ 
&= \min_{D \geq 0} \left[ R(D,\mathbf{X}) + \a D \right],\;\;{\rm a.s.}
\end{align}

\end{remark}

\section{Universal lossy coding via MCMC}\label{sec: universal lossy}

In this section, we will show how simulated annealing Gibbs sampling enables us to get close to the performance of the impractical exhaustive search coding algorithm described in the previous section. Throughout this section we fix the slope $\a > 0$.

Associate with each reconstruction sequence $y^n$ the
\emph{energy}
\begin{align} 
\Ec(y^n) &\triangleq n \left[ H_k (y^n) + \a d_n (x^n, y^n) \right] \nonumber \\
& = \sum\limits_{\mathbf{u}\in\hat{\Xc}^k}\mathbf{1}^T\mb_{\cdot,\mathbf{u}}(y^n)\Hc\left(\mb_{\cdot,\mathbf{u}}(y^n) \right) + \a \sum\limits_{i=1}^n d(x_i,y_i).
\end{align}
The \emph{Boltzmann distribution} can now be defined as the pmf on $\hat{\Xc}^n$ given by
\begin{equation}\label{eq: Boltzmann distribution}
    p_{\beta} (y^n) = \frac{1}{Z_{\beta}} \exp \{ - \beta \Ec(y^n) \},
\end{equation}
where $Z_{\beta}$ is the normalization constant (partition function).
Note that, though this dependence is suppressed in the notation for
simplicity,  $\Ec(y^n)$, and therefore also $p_{\beta}$ and $Z_{\beta}$
depend on $x^n$ and $\a$, which are fixed until further notice. When
$\beta$ is large and $Y^n \sim p_\beta$, then with high probability
\begin{equation}  \label{eq: exhaustive min}
   H_k (Y^n) + \a d_n (x^n, Y^n)\approx \min_{y^n} \left[  H_k (y^n) + \a d_n (x^n,y^n)  \right] .
\end{equation}
Thus, for large $\b$, using a sample from the Boltzmann distribution $p_{\beta}$ as the reconstruction sequence, would yield
performance close to that of an exhaustive search scheme that would use the achiever of the minimum in (\ref{eq: exhaustive min}).
 Unfortunately, it is hard to sample from the Boltzmann distribution directly. We can, however, get approximate samples via MCMC, as we describe next.

As mentioned earlier, the Gibbs sampler \cite{Gibbs_sampler} is useful in cases where one is interested in sampling from a probability
distribution which is hard to compute, but the conditional distribution of each variable given the rest of the variables is accessible. In our case, the conditional probability under $p_\beta$ of $Y_i$ given the other variables $Y^{n \setminus i}\triangleq \{Y_n: n\neq i\}$ can be expressed as
\begin{align}\label{eq: computation of the conditional to begin}
p_{\beta} (Y_i = a | Y^{n \setminus i} = y^{n \setminus i}) & = \frac{p_{\beta} (Y_i = a , Y^{n \setminus i} = y^{n \setminus i})}{\sum\limits_{b\in\hat{\Xc}} p_{\beta} (Y_i = b , Y^{n \setminus i} = y^{n \setminus i})}, \\
    &= \frac{\exp \{- \beta \Ec(y^{i-1}ay_{i+1}^n) \}}{\sum\limits_{b\in\hat{\Xc}} \exp \{ - \beta \Ec(y^{i-1}by_{i+1}^n)\}}, \\
    &= \frac{\exp \{ - \beta n \left[ H_k (y^{i-1}ay_{i+1}^n) + \a d_n (x^n,y^{i-1}ay_{i+1}^n) \right] \}}{\sum\limits_{b\in\hat{\Xc}} \exp \{ - \beta n \left[ H_k (y^{i-1}by_{i+1}^n) + \a d_n (x^n,
    y^{i-1}by_{i+1}^n) \right] \}}, \\
    &= \frac{1}{\sum\limits_{b\in\hat{\Xc}} \exp \{ - \beta  \left[ n\Delta H_k (y^{i-1}by_{i+1}^n,a)  + \a \Delta d(b,a,x_i) \right]\}}, \label{eq: conditional prob of gibbs sampler}
\end{align}
where $\Delta H_k(y^{i-1}by_{i+1}^n,a)$ and $\Delta d(y^{i-1}by_{i+1}^n,a,x_i)$ are defined as
\begin{equation}
\Delta H_k(y^{i-1}by_{i+1}^n,a) \triangleq H_k(y^{i-1}by_{i+1}^n) - H_k(y^{i-1}ay_{i+1}^n),
\end{equation}
and
\[
\Delta d(b,a,x_i) \triangleq d(b,x_i) - d(a,x_i),
\]
respectively.

Evidently, $p_{\beta} (Y_i = y_i | Y^{n \setminus i} = y^{n \setminus i})$ depends on $y^n$ only through
$\{ H_k (y^{i-1}by_{i+1}^n) - H_k(y^{i-1}ay_{i+1}^n) \}_{a,b \in\hat{\Xc}}$ and
$\{d(x_i, a)\}_{a\in\hat{\Xc}}$. In turn, $\{ H_k (y^{i-1}by_{i+1}^n) - H_k (y^{i-1}ay_{i+1}^n) \}_{a,b}$ depends on $y^n$ only through $\{ \mb(y^{i-1}by_{i+1}^n) \}_{b}$.

Note that, given $\mb(y^n)$, the number of operations required to obtain $\mb(y^{i-1}by_{i+1}^n)$, for any $b \in \hat{\Xc}$
is linear in $k$, since the number of contexts whose counts are affected by the change of one component of $y^n$ is at most $2k+2$. To be more specific, letting $\mathcal{S}_i (y^n,b)$ denote the set of contexts whose counts are affected when the $i^{\rm th}$ component of $y^n$ is flipped from $y_i$ to $b$, we have $|\Sc_i (y^n, b)|\leq 2k+2$. Further, since
\begin{align}
&n[H_k(y^{i-1}by_{i+1}^n)- H_k (y^{i-1}ay_{i+1}^n )] =\sum_{\ub \in \mathcal{S}_i (y^{i-1}by_{i+1}, a)} \nonumber\\
&\left[\mathbf{1}^T \mb_{\cdot,\ub} (y^{i-1}by_{i+1}^n)\mathcal{H} \left( \mb_{.\ub} ( y^{i-1}by_{i+1}^n) \right)
-\mathbf{1}^T \mb_{\cdot,\ub}(y^{i-1}ay_{i+1}^n)\mathcal{H} \left(\mb_{\cdot,\ub}(y^{i-1}ay_{i+1}^n) \right)\right],
\end{align}
it follows that, given $\mathbf{m}(y^{i-1}by_{i+1}^n)$ and $H_k (y^{i-1}by_{i+1}^n)$,
the number of operations required to compute $\mathbf{m}(y^{i-1}ay_{i+1}^n)$ and $H_k ( y^{i-1}ay_{i+1}^n )$ is linear in $k$ (and independent of $n$).

Now consider the following algorithm (Algorithm 1 below) based on the Gibbs sampling for sampling from $p_\beta$, and let $\hat{X}^n_{\a,r}(X^n)$ denote its (random) outcome when taking $k=k_n$ and $\mathbf{\beta}=\{\beta_t\}_t$ to be  deterministic sequences satisfying $k_n = o( \log n)$ and $\beta_t = \frac{1}{T_0^{(n)}}\log(\lfloor\frac{t}{n}\rfloor+1)$, for some $T_0^{(n)}>n\Delta$, where
\begin{align}
\Delta=\max_{i}\max\limits_{\scriptsize{\left\{\begin{array}{c}
                             u^{i-1} \in \hat{\Xc}^{i-1}, \\
                             u_{i+1}^n \in \hat{\Xc}^{n-i}, \\
                             a,b \in \hat{\Xc}
                           \end{array}\right.}}
\left|\Ec(u^{i-1}au_{i+1}^n)-\Ec(u^{i-1}bu_{i+1}^n)\right|,
\end{align}
applied to the source sequence $X^n$ as input.\footnote{Here and throughout it is implicit that the randomness used in the algorithms is independent of the source, and the randomization variables used at each drawing are independent of each other.} By the previous discussion, the computational complexity of the algorithm at each iteration is independent of $n$ and linear in $k$.
\begin{theorem} \label{th: the MCMC based scheme}
Let $\mathbf{X}$ be a stationary ergodic source.  Then
\begin{align}\label{eq: achieving optimal point on the rd curve}
\lim_{n\rightarrow \infty} \lim_{r \rightarrow \infty} \left[ \frac{1}{n} \ell_{{\sf\footnotesize  LZ}} \left( \hat{X}^n_{\a,  r} (X^n) \right)  + \a d_n (X^n, \hat{X}^n ) \right] = \min_{D \geq 0} \left[ R(D,\mathbf{X}) + \a D \right], \;\;{\rm a.s}.
\end{align}
\end{theorem}
\begin{proof}
The proof is presented in Appendix A.
\end{proof}

\begin{algorithm}[h!]
\caption{Generating the reconstruction sequence}
\label{alg: mcmc_lossy_coder}
\begin{algorithmic}[1]
\REQUIRE $x^n$, $k$, $\alpha$, $\{ \beta_t \}_t$, $r$
\ENSURE a reconstruction sequence $\hat{x}^n$
\STATE $y^n \leftarrow x^n$
\FOR{$t=1$ to $r$}
\STATE Draw an integer $i \in \{1, \ldots, n \}$ uniformly at random
\STATE For each $b \in \hat{\Xc}$ compute $p_{\beta_t} (Y_i = b | Y^{n \setminus i} = y^{n \setminus i})$ given in (\ref{eq: conditional prob of gibbs sampler})
\STATE Update $y^n$ by replacing its $i^{\rm th}$ component $y_i$ by  $Z$, where \\$Z\sim p_{\beta_t} (Y_i=\cdot | Y^{n \setminus i} = y^{n \setminus i})$
\STATE Update $\mb(y^{n})$ and $H_k (y^n)$
\ENDFOR \STATE $\hat{x}^n \leftarrow y^n$
\end{algorithmic}
\end{algorithm}

\section{Sliding-window rate-distortion coding via \small{MCMC}} \label{sec: sliding}

The classical approach to lossy source coding is block coding initiated by Shannon \cite{Shannon48}. In this method, each possible source block of length $n$ is mapped into a reconstruction block of the same length. One of the disadvantages of this method is that applying a block code to a stationary process converts it into a non-stationary reconstruction process. Another approach to the rate-distortion coding problem is sliding-block (SB), a.s.~stationary, coding introduced by R.M.~Gray, D.L. Neuhoff, and D.S.~Ornstein in \cite{Gray_Neuhoff_Ornstein75} and also independently by K. Marton in \cite{Marton_SB} both in 1975. In this method, a fixed SB map of a certain order $2k_f+1$ slides over the source sequence and generates the reconstruction sequence which has lower entropy rate compared to the original process. The advantage of this method with respect to the block coding technique is that while the achievable rate-distortion regions of the two methods provably coincide, the stationarity of the source is preserved by a SB code \cite{Gray_SB}. Although SB codes seem to be a good alternative to block codes, there has been very little progress in constructing good such codes since their introduction in 1975, and to date there is no known practical method for finding practical SB codes. In this section we show how our MCMC-based approach can be applied to find good entropy-constrained SB codes of a certain order $2k_f+1$.

There are a couple of advantages in using SB codes instead of block codes.  One main benefit is getting rid of the blocking artifacts resulting from applying the code to non-overlappying adjacent blocks of data. This issue has been extensively studied in image compression, and one of the reasons wavelet transform is preferred over more traditional image compression schemes like DCT is that it can be implemented as a sliding-window transform, and therefore does not introduce blocking artifacts \cite{book:Mallat}. The other advantage of SB codes is in terms of speed  and more memory-efficiency. 

\begin{remark} There is a slight difference between SB codes  proposed in \cite{Gray_Neuhoff_Ornstein75}, and our entropy-constrained SB codes. In \cite{Gray_Neuhoff_Ornstein75}, it is assumed that after the encoder converts the source process into the coded process, with no more encryption, it can be directly sent to the decoder via a channel that has capacity of $R$ bits per transmission. Then the decoder, using another SB code, converts the coded process into the reconstruction process. In our setup on the other hand, the encoder directly converts the source process into the reconstruction process, which has lower entropy, and then employs a universal lossless coder to describe the coded sequence to the decoder. The decoder then applies the universal lossless decoder that corresponds to the lossless encoder used at the encoder to retrieve the reconstruction sequence.
\end{remark}
A SB code of window length $2k_f+1$, is a function $\textsl{f}:\mathcal{X}^{2k_f+1}\rightarrow\hat{\Xc}$ which is applied to the source process $\{X_n\}$ to construct the reconstruction block as follows
\begin{equation}
\hat{X}_i=\textsl{f}(X_{i-k_f}^{i+k_f}).
\end{equation}
The total number of $(2k_f+1)$-tuples taking values in $\mathcal{X}$ is
\[
K_f=|\mathcal{X}|^{2k_f+1}.
\]
Therefore, for specifying a SB code of window length $2k_f+1$, there are $K_f$ values to be determined, and $\textsl{f}$ can be represented as a vector $f^{K_f}=[f_0,f_1,\ldots,f_{K_f-1}]$ where $f_i\in\hat{\Xc}$ is the output of function $\textsl{f}$ to the input vector $\textbf{b}$ equal to the expansion of $i$ in $2k_f+1$ symbols modulo $|\mathcal{X}|$, i.e., $i=\sum\limits_{j=0}^{2k_f}b_j|\mathcal{X}|^j$.

For coding a source output sequence $x^n$ by a SB code of order $2k_f+1$, among $|\hat{\Xc}|^{|\mathcal{X}|^{2k_f+1}}$ possible choices, similar to the exhaustive search algorithm described in Section \ref{sec: universal lossy}, here we look for  the one that minimizes the energy function assigned to each possible SB code as
\begin{equation}\label{eq: energy function to SB code}
\Ec(f^{K_f}) \triangleq n\left[H_k(y^n) + \a d_n(x^n,y^n)\right],
\end{equation}
where $y^n = y^n[x^n,f^{K_f}]$ is defined by $y_i=\textsl{f}(x_{i-k_f}^{i+k_f})$. Like before, we consider a cyclic rotation as $x_i=x_{i+n}$, for any $i\in\mathds{N}$. Again, we resort to the simulated annealing Gibbs sampling method in order to find the minimizer of (\ref{eq: energy function to SB code}). Unlike in (\ref{eq: Boltzmann distribution}), instead of the space of possible reconstruction blocks, here we define Boltzmann distribution over the space of possible SB codes. Each SB code is  represented by a unique vector $f^{K_f}$, and  $p_{\beta}(f^{K_f})\propto \exp{(-\beta \Ec(y^n))}$, where $y^n=y^n[x^n,f^{K_f}]$. The conditional probabilities required at each step of the Gibbs sampler can be written as
\begin{align}
p_{\beta}(f_i = \theta|f^{K_f\backslash i}) &= \frac{p_{\beta}(f^{i-1}\theta f_{i+1}^{K_f})}{\sum\limits_{\vartheta}p_{\beta}(f^{i-1}\vartheta f_{i+1}^{K_f})},\\
&= \frac{1}{\sum\limits_{\vartheta}\exp{(-\beta (\Ec(f^{i-1}\vartheta f_{i+1}^{K_f})-\Ec(f^{i-1}\theta f_{i+1}^{K_f})))}}.\label{eq: SB code conditional prob}
\end{align}
Therefore, for computing the conditional probabilities we need to find out by how much changing one entry of $f^{K_f}$ affects the energy function. Compared to the previous section, finding this difference in this case is more convoluted and should be handled with more deliberation. To achieve this goal, we first categorize different positions in $x^n$ into $|\mathcal{X}|^{2k_f+1}$ different types and construct the $s^n$ vector such that the label of $x_i$, $\alpha_i$, is defined to be
\begin{align}
\alpha_i \triangleq\sum\limits_{j=-k_f}^{k_f}x_{n+j}|\mathcal{X}|^{k_f+j}.
\end{align}
In other words, the label of each position is defined to be the symmetric context of length $2k_f+1$ embracing it, i.e., $x_{i-k_f}^{i+k_f}$. Using this definition, applying a SB code $f^{K_f}$ to a sequence $x^n$ can alternatively be expressed as constructing a sequence $y^n$ where
\begin{equation}
y_i = f_{\alpha_i}.
\end{equation}
From this representation, changing $f_i$ from $\theta$ to $\vartheta$ while leaving the other elements of $f^{K_f}$ unchanged only affects the positions of the $y^n$ sequence that correspond to the label $i$ in the $s^n$ sequence, and we can write the difference between energy functions appearing in (\ref{eq: SB code conditional prob}) as
\begin{align}
\Ec(f^{i-1}\vartheta f_{i+1}^{K_f})-\Ec(f^{i-1}\theta f_{i+1}^{K_f}) &= \nonumber \\
&\hspace{-2.5cm}n\left[ H_k(\mb(y^n)-H_k(\mb(\hat{y}^n)\right] + \a \sum\limits_{j:\alpha_j=i}(d(x_j,\vartheta)-d(x_j,\theta)), \label{eq: SB code, delta E}
\end{align}
where $y^n$ and $\hat{y}^n$ represent the results of applying $f^{i-1}\vartheta f_{i+1}^{K_f}$ and $f^{i-1}\theta f_{i+1}^{K_f}$ to $x^n$ respectively, and as noted before the two vectors differ only at the positions $\{j: \alpha_j=i\}$. Flipping each position in the $y^n$ sequence in turn affects at most $2(k+1)$ columns of the count matrix $\mb(y^n)$. Here at each pass of the Gibbs sampler a number of positions in the $y^n$ sequence are flipped simultaneously.
Algorithm \ref{alg: SB_update_count} describes how we can keep track of all these changes and update the count matrix. After that in analogy to
Algorithm  \ref{alg: mcmc_lossy_coder},  Algorithm ~\ref{alg: find_f_mcmc_SB} runs the Gibbs sampling method to find the best SB code of order $2k_f+1$, and at each iteration it employs Algorithm ~ \ref{alg: SB_update_count}.

\begin{algorithm}[h]
\floatstyle{boxed}
\caption{Updating the count matrix of $y^n=\textsl{f}(x^n)$, when $f_i$ changes from $\theta$ to $\vartheta$}
\label{alg: SB_update_count}
\begin{algorithmic}[1]
\REQUIRE $x^n$, $k_f$, $k$, $\mb(y^n)$, $i$, $\vartheta$, $\theta$
\ENSURE $\textbf{m}(\hat{y}^n)$
\STATE $a^n\leftarrow \textbf{0}$
\STATE $\hat{y}^n \leftarrow y^n$
\FOR{$j=1$ to $n$}
   \IF{$\alpha_j = i$}
     \STATE $\hat{y}_j \leftarrow \theta$
   \ENDIF
\ENDFOR
\STATE $\textbf{m}(\hat{y}^n)\leftarrow\mb(y^n)$
\FOR{$j=k_f+1$ to $n-k_f$}
    \IF{$\alpha_j = i$}
        \STATE $a_j^{j+k}\leftarrow\textbf{1}$
    \ENDIF
\ENDFOR

\FOR{$j=k+1$ to $n-k$}
    \IF{$a_j=1$}
        \STATE $m_{y_j,y_{j-k}^{j-1}} \leftarrow m_{y_j,y_{j-k}^{j-1}}-1$
        \STATE  $m_{\hat{y}_j,\hat{y}_{j-k}^{j-1}} \leftarrow m_{\hat{y}_j,\hat{y}_{j-k}^{j-1}}+1$
    \ENDIF
\ENDFOR
\end{algorithmic}
\end{algorithm}

\begin{algorithm}[h]
\caption{Universal SB lossy coder based on simulated annealing Gibbs sampler}
\label{alg: find_f_mcmc_SB}
\begin{algorithmic}[1]
\REQUIRE $x^n$, $k_f$, $k$, $\a$, $\beta$, $r$
\ENSURE $f^{K_f}$
\FOR{$t=1$ to $r$}
    \STATE Draw an integer $i \in \{1, \ldots, K_f\}$ uniformly at random
    \STATE For each $a \in \hat{\Xc}$ compute $p_{\beta_t} (f_i = \theta | f^{K_f \setminus i} )$ using Algorithm \ref{alg: SB_update_count}, equations
    (\ref{eq: SB code conditional prob}), and (\ref{eq: SB code, delta E})
    \STATE Update $f^{K_f}$ by replacing its $i^{\rm th}$ component $f_i$ by  $\theta$ drawn from the pmf computed in the previous step
\ENDFOR
\end{algorithmic}
\end{algorithm}
Let $f^{K_f^{(n)}}_{\beta,\a,r}$ denote the output of Algorithm ~\ref{alg: find_f_mcmc_SB} to input vector $x^n$ at slope $\a$ after $r$ iterations, and annealing process $\beta$. $K_f^{(n)}=2^{2k_f^{(n)}+1}$ denotes the length of the vector $f$ representing the SB code. The following theorem proved in Appendix B states that Algorithm ~\ref{alg: find_f_mcmc_SB} is asymptotically optimal for any stationary ergodic source; i.e.,coding a source sequence by applying the SB code $f^{K_f^{(n)}}_{\beta,\a,r}$ to the source sequence, and then describing the output to the decoder using Lempel-Ziv algorithm, asymptotically, as the number of iterations and window length $k_f$ grow to infinity, achieves the rate-distortion curve.

\begin{theorem} \label{th: the MCMC based scheme for finding SB code}
Given a sequence $\{k_f^{(n)}\}$ such that $k_f^{(n)}\rightarrow\infty$, schedule $\beta_t^{(n)}=\frac{1}{T_0^{(n)}}\log(\lfloor \frac{t}{K_f^{(n)}}\rfloor+1)$ for some $T_0^{(n)}>K_f\Delta$, where
\begin{align}\label{eq: def Delta SB code}
\Delta=\max_{i}\max\limits_{\scriptsize{\left\{\begin{array}{c}
                             f^{i-1} \in \hat{\Xc}^{i-1}, \\
                             f_{i+1}^n \in \hat{\Xc}^{K_f-i}, \\
                             \vartheta,\theta \in \hat{\Xc}
                        \end{array}\right.}}
|\Ec(f^{i-1}\vartheta f_{i+1}^{K_f})-\Ec(f^{i-1}bf_{i+1}^{K_f})|,
\end{align}
and $k=o(\log n)$. Then, for any stationary ergodic source $\mathbf{X}$, we have
\begin{align}\label{eq: achieving optimal point on the rd curve by SB code}
\lim_{n\rightarrow \infty}\lim_{r \rightarrow \infty} \E \left[ \frac{1}{n} \ell_{{\sf\footnotesize  LZ}} \left(\hat{X}^n \right) + \a d_n(X^n, \hat{X}^n) \right] = \min_{D \geq 0} \left[R(D,\mathbf{X})+ \a D\right],
\end{align}
where $\hat{X}^n$ is the result of applying SB code $f^{K_f}_{\beta,\a,r}$ to $X^n$.
\end{theorem}
\begin{proof}
The proof is presented in Appendix B.
\end{proof}
Note that in Algorithm ~\ref{alg: find_f_mcmc_SB}, for a fixed $k_f$, the SB code is a vector of length $K_f=|\Xc|^{2k_f+1}$. Hence, the size of the search space is $|\hat{\Xc}|^{K_f}$ which is independent of $n$. Moreover, the transition probabilities of the SA as defined by (\ref{eq: SB code conditional prob}) depend on the differences of the form presented in (\ref{eq: SB code, delta E}), which, for a stationary ergodic source and fixed $k_f$, if $n$ is large enough, linearly scales with $n$. I.e., for a given $f^{i-1}$, $f_{i+1}^{K_f}$, $\vartheta$ and $\theta$,
\begin{align}
\lim\limits_{n\rightarrow\infty}\frac{1}{n}[\Ec(f^{i-1}\vartheta f_{i+1}^{K_f})-\Ec(f^{i-1}\theta f_{i+1}^{K_f})]=q\quad\quad\textmd{a.s.},
\end{align}
where $q\in[0,1]$ is some fixed value depending only on the source distribution. This is an immediate consequence of the ergodicity of the source plus the fact that SB coding of a stationary ergodic process results in another process which is jointly stationary with the initial process and is also ergodic. On the other hand, similar reasoning proves that $\Delta$ defined in (\ref{eq: def Delta SB code}) scales linearly by $n$. Therefore, overall, combining these two observations, for large values of $n$ and fixed $k_f$, the transition probabilities of the nonhomogeneous MC defined by the SA algorithm incorporated in Algorithm ~\ref{alg: find_f_mcmc_SB} are independent of $n$. This does not mean that the convergence rate of the algorithm is independent of $n$, because for achieving the rate-distortion function one needs to increase $k_f$ and $n$ simultaneously to infinity.

\section{Simulation results}\label{sec: simulations}

We dedicate this section to the presentation of some initial experimental results obtained by applying the schemes presented in the previous sections on simulated and real data. The Sub-section \ref{sec: sub1} demonstrates the performance of  Alg.~\ref{alg: mcmc_lossy_coder} on simulated 1-D and real 2-D data. Some results on the application Alg.~\ref{alg: find_f_mcmc_SB} on simulated 1D data is shown in Sub-section \ref{sec: sub2}.

\subsection{Block coding}\label{sec: sub1}

In this sub-section, some of the simulation results obtained from applying Alg.~\ref{alg: mcmc_lossy_coder} of Section \ref{sec: universal lossy} to real and simulated data are presented. The algorithm is easy to apply, as is, to both 1-D and 2-D data .

\begin{figure}
\begin{center}
\includegraphics[height=9cm,width=10cm]{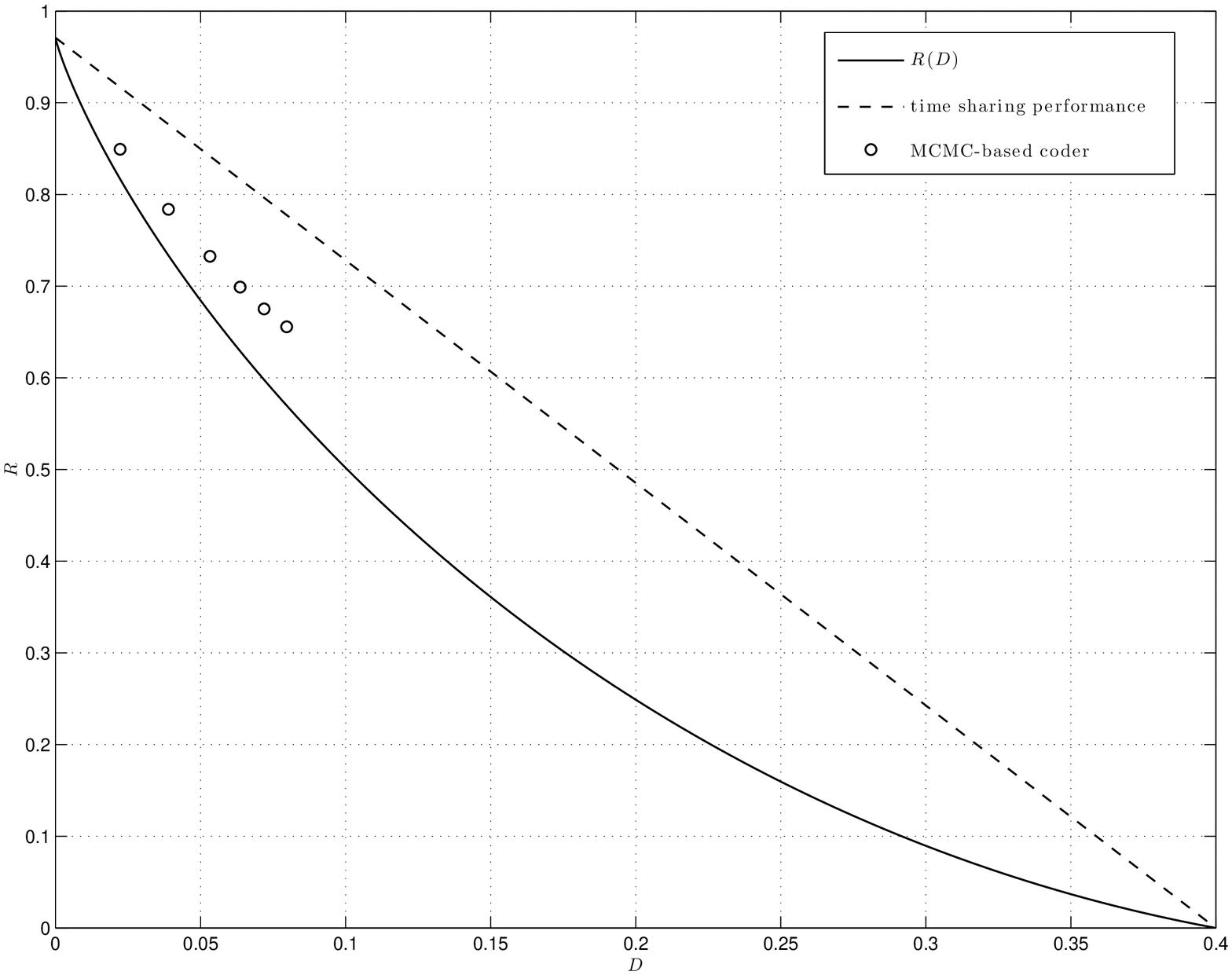}
\caption{Comparing the Alg.~\ref{alg: mcmc_lossy_coder} performance with the optimal rate-distortion tradeoff for a $\Bern(p)$ i.i.d.~source, $p=0.4$ ($n=15\times10^3$, $k=9$, $\beta_t = (1/\gamma)^{\lceil t/n \rceil}$, $\gamma=0.75$, $r=10n$ and $\alpha=4:-0.4:2$).}
\label{fig: iid 0.4 source}
\end{center}
\end{figure}

As the first example, consider a $\Bern(p)$ i.i.d~source. Fig.~\ref{fig: iid 0.4 source} compares the optimal rate-distortion tradeoff against Alg.~\ref{alg: mcmc_lossy_coder} performance for  $p=0.4$ respectively. The algorithm parameters  are $n=15\times 10^3$, $k=9$, $\beta_t = (1/\gamma)^{\lceil t/n \rceil}$, where $\gamma=0.75$, and $\alpha=4:-0.4:2$.   Each point  corresponds to the average performance over $N=50$ iterations. At each iteration the algorithm starts from $\a=4$, and gradually decreases the coefficient by $0.4$ at each step. Moreover, except for $\a=4$ where $\hat{x}^n$ is initialized by $x^n$, for each other value of $\alpha$, the algorithm starts from the quantized sequence found for the previous value of $\alpha$.

\begin{figure}
\begin{center}
\includegraphics[height=9cm,width=10cm]{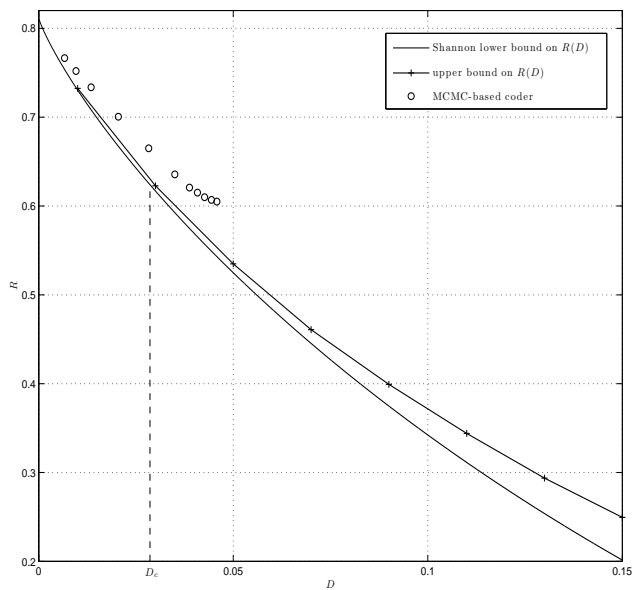}
\caption{Comparing the algorithm rate-distortion performance with  Shannon lower bound for a BSMS($p$) ($p=0.2$, $n=2\times10^4$, $k=8$, $\beta_t = (1/\gamma)^{\lceil t/n \rceil}$, $\gamma=0.8$, $r=10n$ and $\alpha=5:-0.5:3$) }
\label{fig: bsms RD}
\end{center}
\end{figure}

As another example, Fig.~\ref{fig: bsms RD} compares the performance of Alg.~\ref{alg: mcmc_lossy_coder}  when applied to a binary symmetric Markov source (BSMS) with transition probability $p=0.25$ against the Shannon lower bound (SLB)  which sates that for a BSMS
\begin{equation}
R(D)\geq R_{\textmd{SLB}}(D)\triangleq h(p)-h(D).
\end{equation}
There is no known explicit characterization of the rate-distortion tradeoff for a BSMS except for a low distortion region. It has been proven that for $D<D_c$, where
\begin{equation}
D_c=\frac{1}{2}\left(1-\sqrt{1-(p/q)^2}\right),
\end{equation}
the SLB holds with equality, and for $D>D_c$, we have strict inequality, i.e.~$R(D)>R_{\textmd{SLB}}$ \cite{Gray_markov_source}. In our case
$D_c= 0.0286$ which is indicated in the figure. For distortions beyond $D_c$, an  upper bound on the rate-distortion function, derived based on the results presented in \cite{ISIT_JW}, is  shown for comparison. The parameters here are: $n=2\times 10^4$, $k=8$, $\beta_t = (1/\gamma)^{\lceil t/n \rceil}$, $\gamma=0.8$, $r=10n$ and $\alpha=5:-0.5:3$.

\begin{figure}
\begin{center}
\includegraphics[height=10cm,width=10cm]{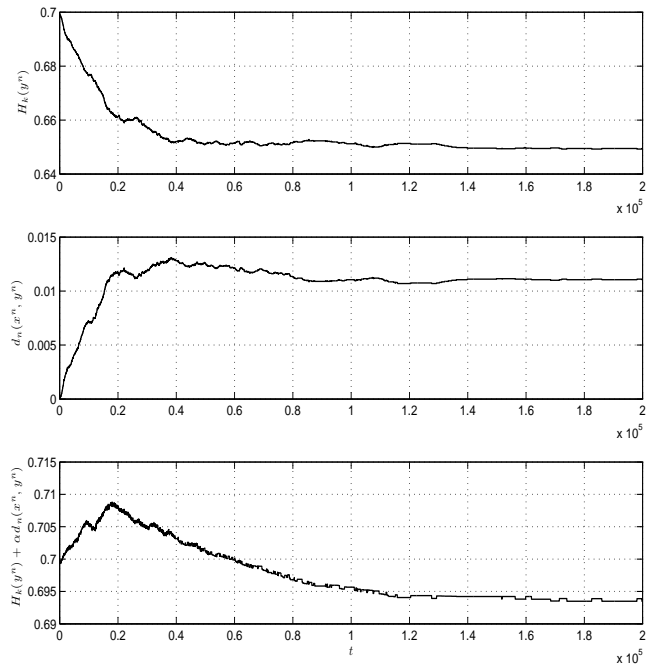}
\caption{Sample paths demonstrating evolutions of the
empirical conditional entropy, average distortion, and energy
function when Alg. \ref{alg: mcmc_lossy_coder} is applied to the output of a $\Bern(p)$ i.i.d~source, with $p=0.2$ ($n=2\times 10^4$, $k=9$, $\alpha=4$, $r=10n$, and $\beta_t=(1/\gamma)^{\lceil t/n \rceil}$, where $\gamma = 0.7$ ).}
\label{fig:RDC1}
\end{center}
\end{figure}

\begin{figure}
\begin{center}
\includegraphics[height=10cm,width=10cm]{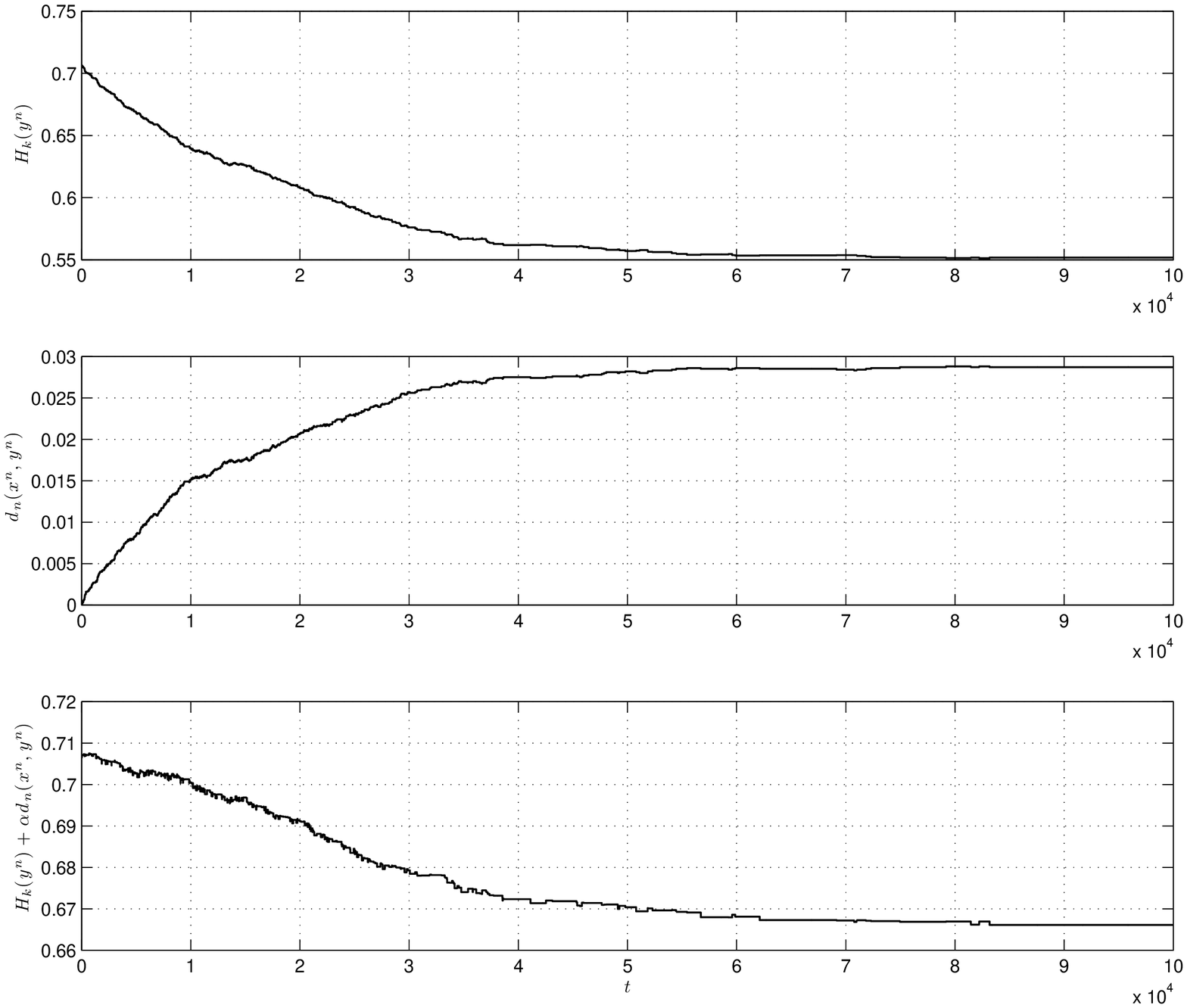}
\caption{Sample paths demonstrating evolutions of the
empirical conditional entropy, average distortion, and energy
function when Alg. \ref{alg: mcmc_lossy_coder} is applied to the output of a BSMS source, with $p=0.2$ ($n= 10^4$, $k=7$, $\alpha=4$, $r=10n$, and $\beta_t=(1/\gamma)^{\lceil t/n \rceil}$, where $\gamma = 0.8$ ).}
\label{fig:RDC2}
\end{center}
\end{figure}

To illustrate the encoding process, Fig.~\ref{fig:RDC1} depicts the evolutions of $H_k(\hat{x}^n)$, $d_n(x^n,\hat{x}^n)$, and $\Ec(\hat{x}^n)=H_k(\hat{x}^n)+\alpha d_n(x^n,\hat{x}^n)$ during coding iterations. It can be observed that, as time proceeds, while the complexity of the sequence has an overall decreasing trend, as expected, its distance with the original sequence increases. The over cost which we are trying to minimize, increases initially, and starts a decreasing trend after a while. Here the source again is a $\Bern(p)$ source, but with $p=0.2$. The algorithm parameters are $n=2\times 10^4$, $k=9$, $\alpha=4$, $r=10n$, and $\beta_t=(1/\gamma)^{\lceil t/n \rceil}$, where $\gamma = 0.7$. Fig.~\ref{fig:RDC2} shows similar curves when the source is binary Markov with transition probability $p=0.2$. The other parameters are  $n= 10^4$, $k=7$, $\alpha=4$, $r=10n$, and $\beta_t=(1/\gamma)^{\lceil t/n \rceil}$, where $\gamma = 0.8$.


\begin{figure}[h]
\begin{center}
\includegraphics[width=25mm]{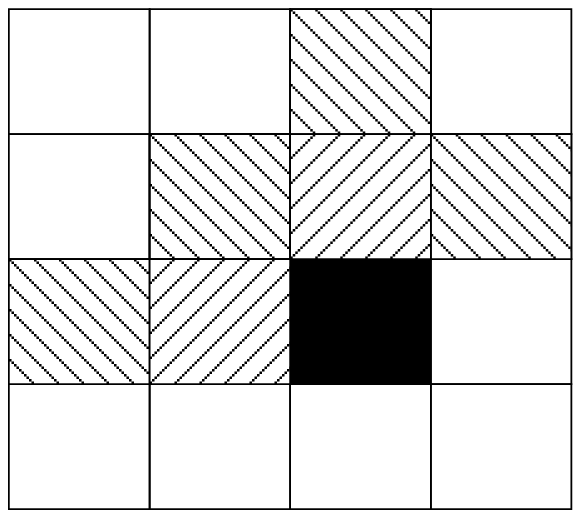}
\caption{The $6^{\rm th}$ order context used in coding of 2-D images}
\label{fig: grid1}
\end{center}
\end{figure}

\begin{figure}
  \begin{center}
	\includegraphics[width=8cm]{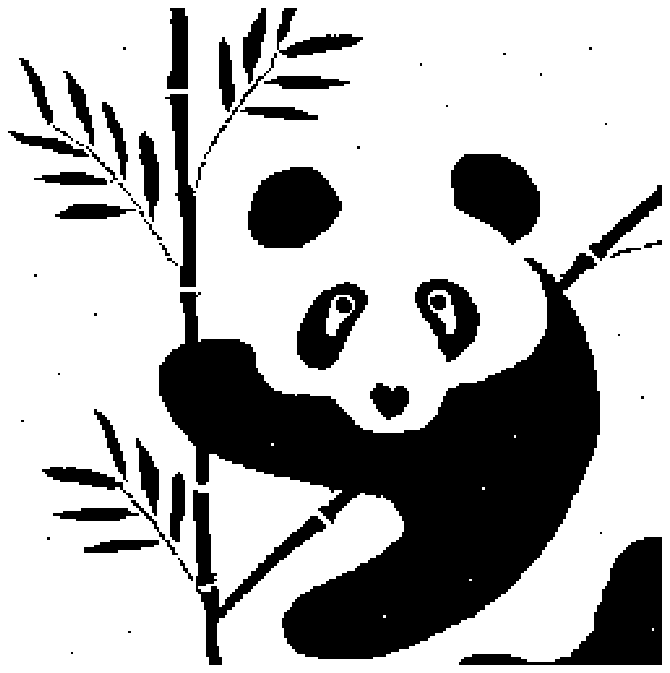} 
    \caption{Original image with empirical conditional entropy of 0.1025}\label{fig: a}
  \end{center}
\end{figure}

\begin{figure}
  \begin{center}
    \subfigure[Reconstruction image with empirical conditional entropy of 0.0600 and average distortion of 0.0337 per pixel ($\alpha=0.1$).] 
    {\label{fig: b} \includegraphics[width=8cm]{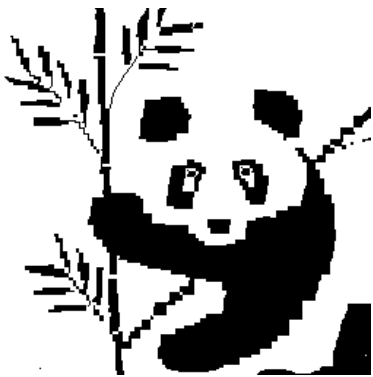} }
    \subfigure[Reconstruction image with empirical conditional entropy of 0.0824 and average distortion of 0.0034 per pixel 
    ($\a= 3.3$).] 
    {\label{fig: c} \includegraphics[width=8cm]{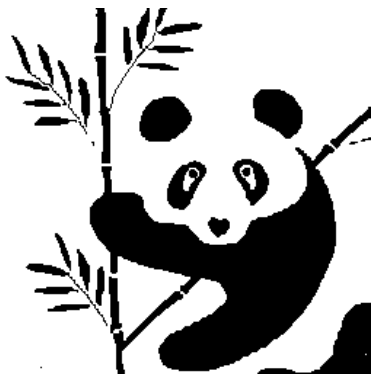} } 
    \caption{Applying Alg.~\ref{alg: mcmc_lossy_coder} to a 2-D binary image ($\beta_t=(1/\gamma)^{\lceil t/n \rceil}$, where $\gamma = 0.99$, $r=50 N$)}
  \end{center}
\end{figure}

Finally, consider applying the algorithm to the $n\times n$ binary image shown in Fig.~\ref{fig: a} , where $n=252$. Let $N\triangleq n^2$ denote the total number of pixels in the image.  Fig.~\ref{fig: b} and Fig.~\ref{fig: c} show the coded version after $r=50N$ iterations for $\alpha = 0.1$ and $\alpha=3.3$ respectively. The algorithm's cooling process is $\beta_t=(1/\gamma)^{\lceil t/n \rceil}$ with $\gamma = 0.99$.  Fig.\ref{fig: grid1} shows the 2-D context used for constructing the count matrix of the image that is used by the algorithm. In the figure, the solid black square represents the location of the current pixel, and the other marked squares denote its $6^{\rm th}$ order causal context that are taken into account. 

\begin{figure}
  \begin{center}
   \includegraphics[width=11cm]{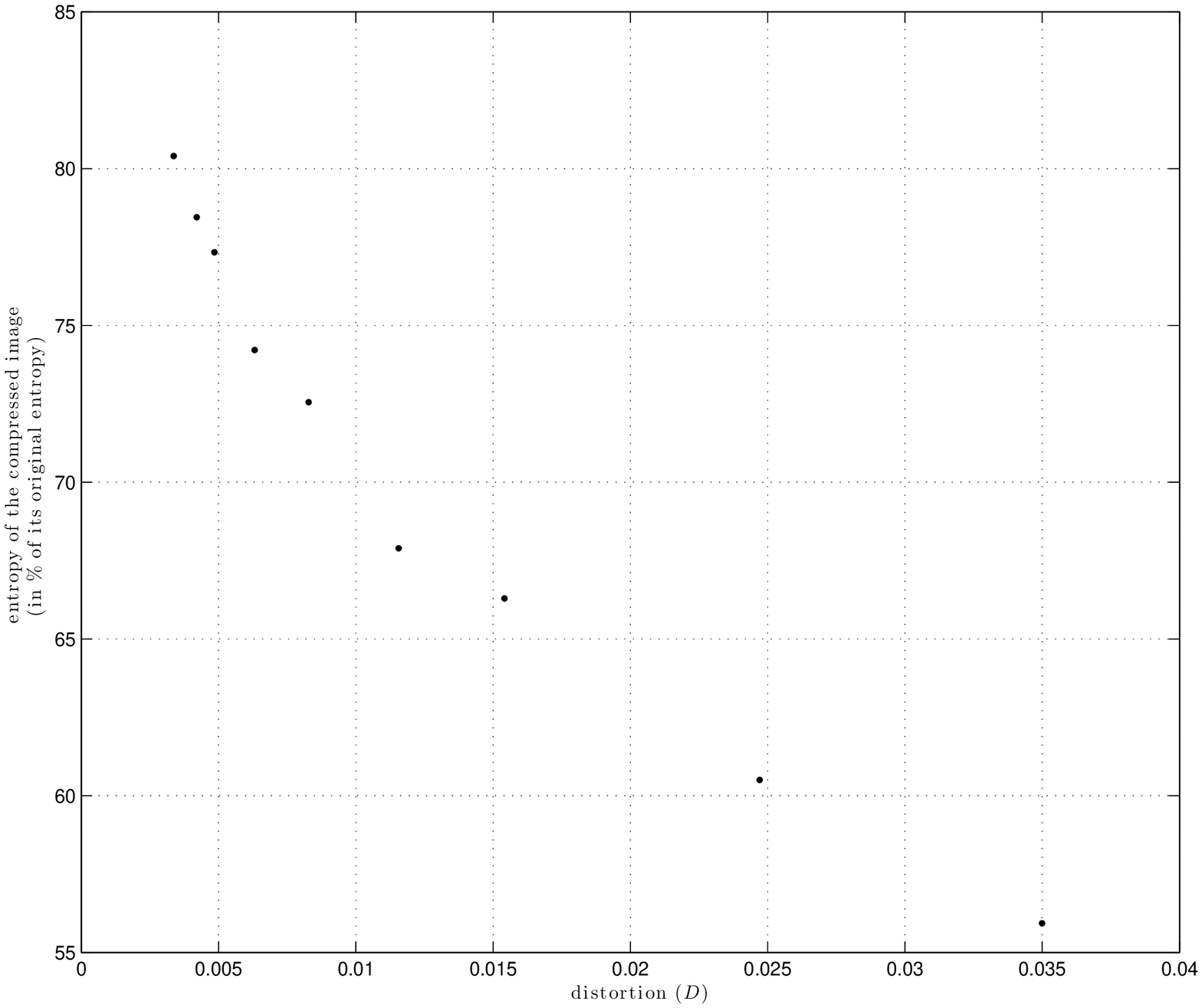}
   \caption{Size of the compressed image in terms of the entropy of the original image (in precentage) versus distortion ($\a=0.1:0.4:3.3$, $\beta_t=(1/\gamma)^{\lceil t/n \rceil}$, where $\gamma = 0.99$, $r=50 N$).}\label{fig: R_D_2dimage}
   \end{center}
\end{figure}

Fig.~\ref{fig: b}, the empirical conditional entropy of the image has decreased from $0.1025$ to $0.0600$ in the reconstruction image, while an average distortion of $D= 0.0337$ per pixel is introduced. Comparing the required space for storing the original image as a PNG file with the amount required for the coded image reveals that in fact the algorithm not only has reduced the conditional empirical entropy of the image by $41.5\%$, but also has cut the size of the file by around $39\%$. Fig.~\ref{fig: R_D_2dimage} shows the size of the compressed image in terms of the size of the original image when $\a$ varies as $0.1:0.4:3.3$.

\subsection{Discussion on  the choice of different parameters}
\subsubsection{Context length $k$} As stated in Theorem \ref{th: the exhaustive search scheme}, in order to get to the optimal performance, $k$ should increase as $o(\log n)$. For getting good performance, it is crucial to choose $k$ appropriately.  Note that the order $k$ determines the order of the count matrix $\mb$ which is used to measure the \emph{complexity} of the quantized sequence. Choosing $k$ to be too big or too small compared to the length of our sequence are both problematic. if $k$ is too small, then the count matrix $\mb$ will not  capture  all useful structures  existing in the sequence. These structures potentially help the universal lossy coder to describe the sequence with fewer number of bits. On the other hand, if $k$ is too large compared to the block length $n$, then $H_k(y^n)$ gives a unreliable underestimate of the complexity of the sequence. One reason is that in this the counts are mainly 0 or some small integer.  

\begin{figure}[h]
  \begin{center}
   \includegraphics[width=11cm]{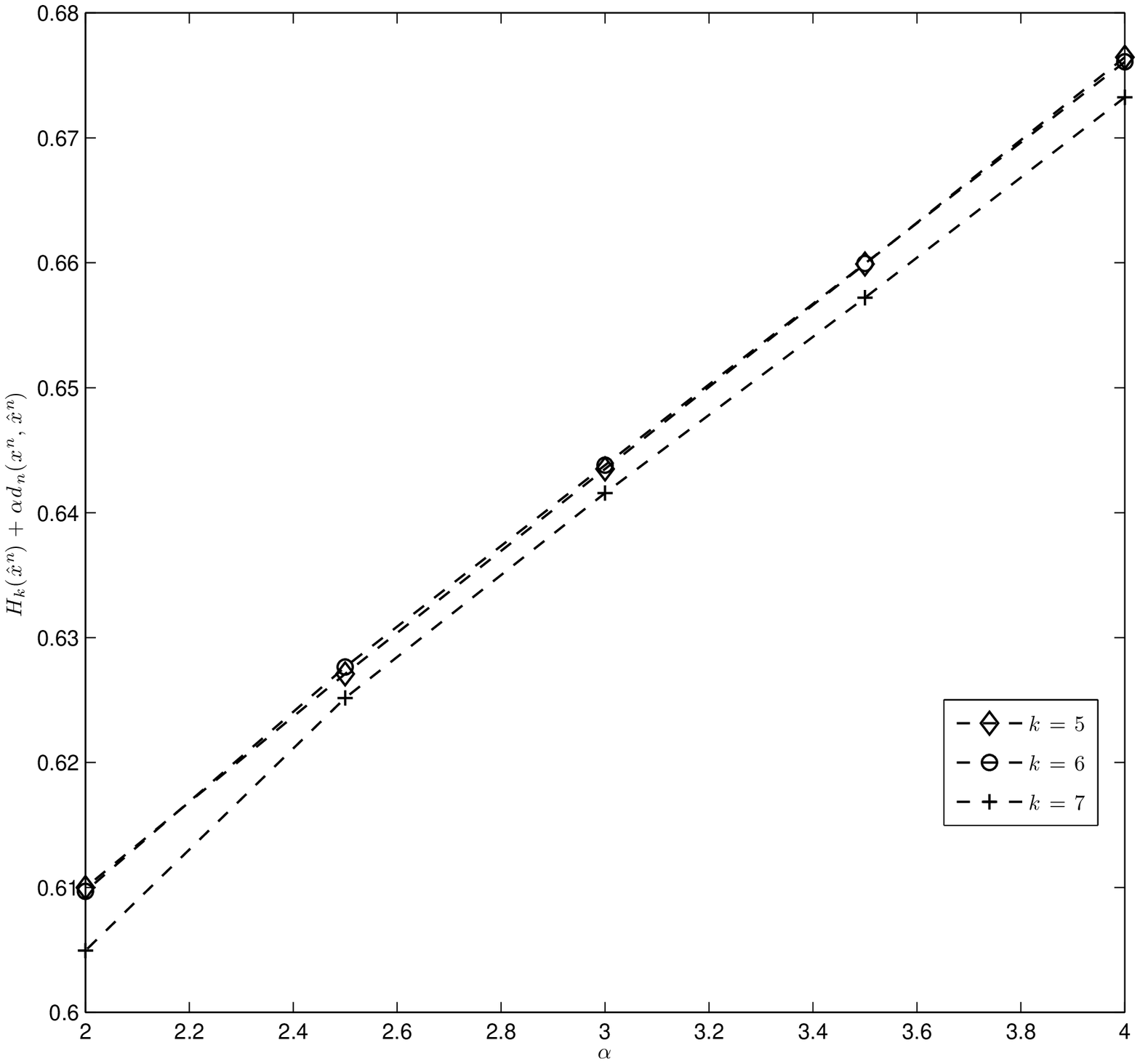}
   \caption{Effect of context length on the algorithm performance in coding a BSMS($p$)  for different values of $\alpha$ ($p=0.2$, $\a=4:-0.5:2$, $\beta_t=(1/\gamma)^{\lceil t/n \rceil}$, where $\gamma = 0.75$, and $r=10 n$).}\label{fig: C_k_bsms}
   \end{center}
\end{figure}

In order to demonstrate the effect of the context length $k$ on the Algorithm performance, consider applying Alg.~\ref{alg: mcmc_lossy_coder} to a binary symmetric Markov source with transition probability $p=0.2$. Fig.~\ref{fig: C_k_bsms} shows the average performance over $I=50$ iterations. The performance measure used in this figure is the average energy of the compressed sequences, i.e., $\Ec(\hat{x}^n)=H_k(\hat{x}^n)+\a d_n(x^n,\hat{x}^n)$, for different values of $\a$. It can be observed that $k=5$ and $k=6$ have almost similar performances, but increasing $k$ to 7 improves the performance noticeably. In each iteration a BSMS($p$) sequence of length $n=10^4$ is generated, and is coded by Alg.~\ref{alg: mcmc_lossy_coder}  for $k=5$, $k=6$ and $k=7$. In all cases the cooling schedule is fixed to $\beta_t=(1/\gamma)^{\lceil t/n \rceil}$, where $\gamma = 0.75$. For each value of $k$, and each simulated sequence, the algorithm starts from $\a=4$ and step by step decreases it  to $2$.

\subsubsection{Block length $n$}
\begin{figure}[h]
  \begin{center}
   \includegraphics[width=11cm]{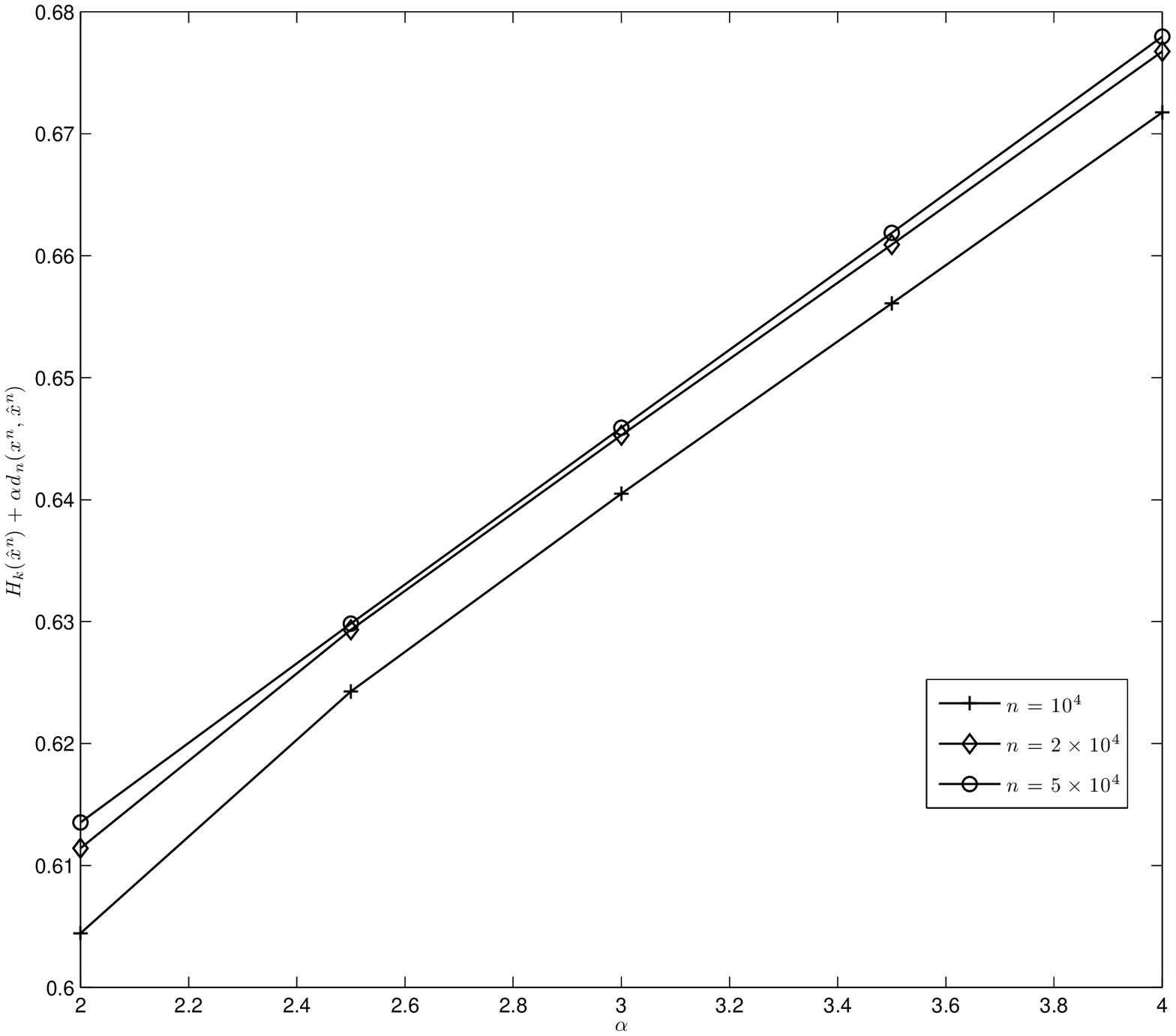}
   \caption{Effect of block length $n$ on the algorithm performance in coding a BSMS($p$)  for different values of $\alpha$ ($p=0.2$,  $k=7$, $\a=4:-0.5:2$, $\beta_t=(1/\gamma)^{\lceil t/n \rceil}$, where $\gamma = 0.75$, and $r=10 n$).}
   \label{fig: C_n_bsms}
   \end{center}
\end{figure}

Fig.~\ref{fig: C_n_bsms} shows the effect of increasing the block length on the minimized cost function for a fixed $k$. The source is again BSMS($p$) with $p=0.2$. The other parameters are $k=7$, $\a=4:-0.5:2$, $\beta_t=(1/\gamma)^{\lceil t/n \rceil}$ with $\gamma = 0.75$, and $r=10 n$. Here, each point corresponds to the average performance over $I=50$ iterations. It can be observed that somewhat counter-intuitively, increasing the block length increases the minimized cost. The reason is that as mentioned earlier, the real cost is not $H_k(\hat{x}^n)+\alpha d_n(x^n,\hat{x}^n)$, but is $\ell_{\rm LZ}(\hat{x}^n)/n+\alpha d_n(x^n,\hat{x}^n)$. This increase in the cost is an indication of the fact that $H_k(\cdot)$ underestimates $\ell_{\rm LZ}(\hat{x}^n)/n$. As $n$ increases the gap between $H_k(\cdot)$ and $\ell_{\rm LZ}(\hat{x}^n)/n$ closes, and the estimate becomes more accurate. Note that while increasing $n$ from $10^4$ increases the cost  noticeably, but from $n=2\times 10^4$ to $n=5\times 10^4$ the increase is almost negligible which somehow suggests that increasing $n$ further will not improve the performance, and for achieving better performance we need to increase $k$ as well as $n$.

\subsubsection{Cooling schedule $\{\b_t\}$}
In all of our simulations the cooling schedule follows the generic form of $\beta_t=\beta_0(1/\gamma)^{\lceil t/n \rceil}$, for some $\gamma<1$, but usually $>0.7$. This is a common schedule used in simulated annealing literature. By this scheme, the running time is divided into intervals of length $n$, and the temperature remains constant during each interval, and decreases by a factor $\gamma$ in the next interval. Hence larger values of $\gamma$ correspond to slower cooling procedures. The specific values of $\gamma$ and $\beta_0$ can be chosen based on the signal to be coded.

\subsubsection{Number of iterations $r$} Although we have not yet derived a convergence rate for Alg.~\ref{alg: mcmc_lossy_coder},  from our simulations results, we suspect that for natural signals not having strange characteristics, $r=mn$ iterations, where $m=o(\log n)$,  is enough for deriving a reasonable approximation of the solution to the exhaustive search algorithm. However, we do not expect similar result to hold for all signals, and there might exist sequences such that the convergence rate of simulated annealing is too slow for them.


\subsection{Sliding-block coding} \label{sec: sub2}

\begin{figure}
\begin{center}
\includegraphics[width=100mm]{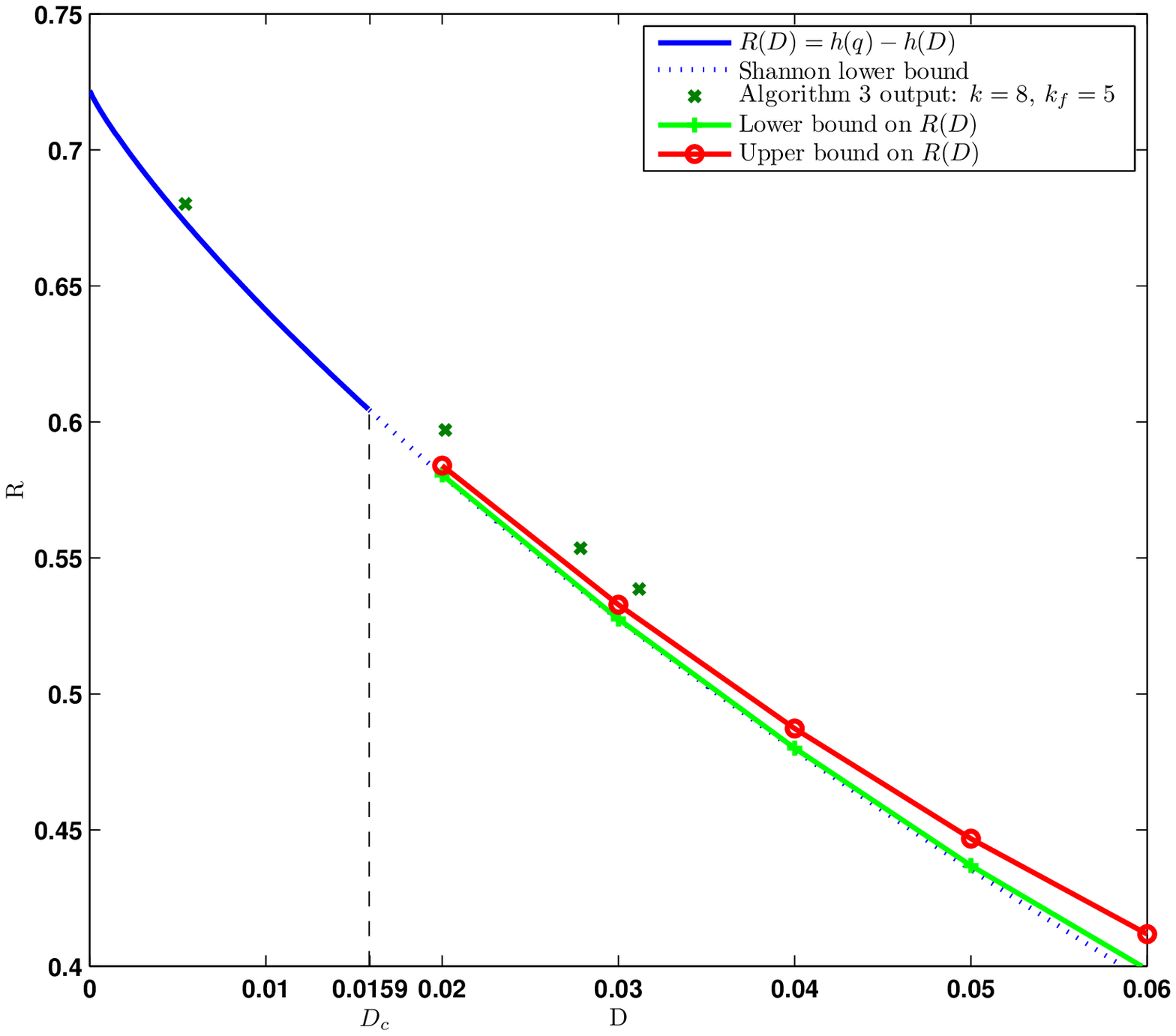}
\caption{Comparing the algorithm rate-distortion performance with the Shannon lower bound for a BSMS with $q=0.2$. Algorithm parameters: $n=5\times10^4$, $k=8$, $k_f=5$ ($K_f=2^{11}$), $\beta_t=K_f\a\log(t+1)$, and slope values $\a=5.25$, $5$, $4.75$ and $4.5$.}
\label{fig: SB code to BSMS(0.2)}
\end{center}
\end{figure}

Consider applying Alg.~\ref{alg: find_f_mcmc_SB} of Section \ref{sec: sliding} to the output of a BSMS with $q=0.2$. Fig.~\ref{fig: SB code to BSMS(0.2)} shows the algorithm output along with Shannon lower bound and lower/upper bounds on $R(D)$ from \cite{ISIT_JW}. Here the parameters are: $n=5\times10^4$, $k=8$, SB window length of $k_f=11$ and $\beta_t=K_f\alpha\log(t+1)$.

In all of the presented simulation results, it is the empirical conditional entropy of the final reconstruction block that we are comparing to the rate-distortion curve. It should be noted that, though this difference vanishes as the block size grows, for finite values of $n$ there would be an extra (model) cost for losslessly describing the reconstruction block to the decoder. 
%

\section{Application: Optimal denoising via MCMC-based lossy coding}\label{sec: denoising}

Consider the problem of denoising a stationary ergodic source $\mathbf{X}$ with unknown distribution corrupted by additive white noise $\mathbf{V}$. Compression-based denoising algorithms have been proposed before by a number of researchers, cf. \cite{Natarajan}, \cite{kolmogrov_sampler}, \cite{Tsachy-Erik-IT-05}  and references therein. The idea of using a universal lossy compressor for denoising was proposed in \cite{kolmogrov_sampler}, and then refined in \cite{Tsachy-Erik-IT-05} to result in a universal denoising algorithm. In this section, we show how our new MCMC-based lossy encoder enables the denoising algorithm proposed in \cite{Tsachy-Erik-IT-05} to lead to an implementable universal denoiser.

In \cite{Tsachy-Erik-IT-05}, it is shown how a universally optimal lossy coder tuned to the right distortion measure and distortion level combined with some simple ``post-processing'' results in a universally optimal denoiser. In what follows we first briefly go over this compression-based denoiser described in \cite{Tsachy-Erik-IT-05}, and then show how our lossy coder can be embedded in for performing the lossy compression part.

Throughout this section we assume that the source, noise, and reconstruction alphabets are $\mathcal{M}$-ary alphabet $\mathcal{A} =\{0,1,\ldots,M-1\}$, and the noise is  additive modulo-$M$ and $P_V(a)>0$ for any $a\in\mathcal{A}$, i.e.~$Z_i=X_i+V_i$.

As mentioned earlier, in the denoising scheme outlined in \cite{Tsachy-Erik-IT-05}, first the denoiser lossily compresses' the noisy signal appropriately, and partly removes the additive noise. Consider a sequence of {\it good} lossy coders characterized by encoder/decoder pairs $(\textrm{E}_n,\textrm{D}_n)$ of block length $n$ working at distortion level $H(V)$ under the difference distortion measure defined as
\begin{equation}\label{eq: definition of difference distortion}
\rho(x,y)=\log\frac{1}{P_V(x-y)}.
\end{equation}
By {\it good}, it is meant that for any stationary ergodic source $\mathbf{X}$, as $n$ grows, the rate distortion performance of the sequence of codes converges to a point on the rate-distortion curve. The next step is a simple ``post-processing'' as follows. For a fixed $m$, define the following count vector over the noisy signal $Z^n$ and its quantized version $Y^n=\textrm{D}_n(\textrm{E}_n(Z^n))$,
\begin{align}
&\hat{Q}^{2m+1}[Z^n,Y^n](z^{2m+1},y) \triangleq \nonumber\\
&\hspace{15mm}\frac{1}{n}|\{1 \leq i \leq n: (Z_{i-k}^{i+k},Y_i)=(z^{2m+1},y)\}|.\label{eq: Q^{2m+1}}
\end{align}
After constructing these count vectors, the denoiser output is generated through the ``post-processing'' or ``de-randomization'' process  as follows
\begin{align}
\hat{X}_i = \argmin\limits_{\hat{x}\in\mathcal{A}} \sum\limits_{y\in\mathcal{A}}\hat{Q}^{2m+1}[Z^n,Y^n](z^{2m+1},y)d(\hat{x},y),
\end{align}
where $d(\cdot,\cdot)$ is the original loss function under which the performance of the denoiser is to be measured. The described denoiser is shown to be universally optimal \cite{Tsachy-Erik-IT-05}, and the basic theoretical justification of this is that the rate-distortion function of the noisy signal $\mathbf{Z}$ under the difference distortion measure satisfies the Shannon lower bound with equality, and it is proved in \cite{Tsachy-Erik-IT-05} that for such sources \footnote{In fact it is shown in \cite{Tsachy-Erik-IT-05} that this is true for a large class of sources including i.i.d sources and those satisfying the Shannon lower bound with equality.} for a fixed $k$, the $k$-th order empirical joint distribution between the source and reconstructed blocks defined as
\begin{align}
&\hat{Q}^k[X^n,Y^n](x^k,y^k) \triangleq \\
&\hspace{15mm}\frac{1}{n}|\{1 \leq i \leq n: (X_i^{i+k-1},Y_i^{i+k-1})=(x^k,y^k)\}|,\nonumber
\end{align}
resulting from a sequence of {\it good} codes converge to $P_{X^k,Y^k}$ in distribution, i.e.~ $\hat{Q}^k[X^n,Y^n]\stackrel{d}{\Rightarrow}P_{X^k,Y^k}$, where $P_{X^k,Y^k}$ is the unique joint distribution that achieves the $k$-th order rate-distortion function of the source. In the case of quantizing the noisy signal under the distortion measure defined in (\ref{eq: definition of difference distortion}), at level $H(V)$, $P_{X^k,Y^k}$ is the $k$-th order joint distribution between the source and noisy signal. Hence, the count vector $\hat{Q}^{2m+1}[Z^n,Y^n](z^{2m+1},y)$ defined in (\ref{eq: Q^{2m+1}}) asymptotically converges to $P_{X_i|Z^n}$ which is what the optimal denoiser would base its decision on. After estimating $P_{X_i|Z^n}$, the post-processing step is just making the optimal Bayesian decision at each position.

The main ingredient of the described denoiser is a universal lossy compressor. Note that the MCMC-based lossy compressor described in Section V is applicable to any distortion measure. The main problem is choosing the parameter $\alpha$ corresponding to the distortion level of interest. To find the right slope, we run the quantization MCMC-based part of the algorithm independently from two different initial points $\alpha_1$ and $\alpha_2$. After convergence of the two runs we compute the average distortion between the noisy signal and its quantized versions. Then assuming a linear approximation, we find the value of $\alpha$ that would have resulted in the desired distortion, and then run the algorithm again from this starting point, and again computed the average distortion, and then find a better estimate of $\alpha$ from the observations so far. After a few repetitions of this process, we have a reasonable estimate of the desired $\alpha$. Note that for finding $\alpha$ it is not necessary to work with the whole noisy signal, and one can consider only a long enough section of data first, and find $\alpha$ from it, and then run the MCMC-based denoising algorithm on the whole noisy signal with the estimated parameter $\alpha$. The outlined method for finding $\alpha$ is similar to what is done in \cite{RV:slope} for finding appropriate Lagrange multiplier.

\subsection{Experiments}

In this section we compare the performance of the proposed denoising algorithm against discrete universal denoiser, {\sf\footnotesize  DUDE} \cite{Erik_seroussi_verdu_Weinberger_tsachy}, introduced in \cite{DUDE}. {\sf\footnotesize  DUDE} is a practical universal algorithm that asymptotically achieves the performance attainable by the best $n$-block denoiser for any stationary ergodic source. The setting of operation of {\sf\footnotesize  DUDE} is more general than what is described in the previous section, and in fact in   {\sf\footnotesize  DUDE} the additive white noise can be replaced by any known discrete memoryless channel.

\begin{figure}
\begin{center}
\includegraphics[width=100mm]{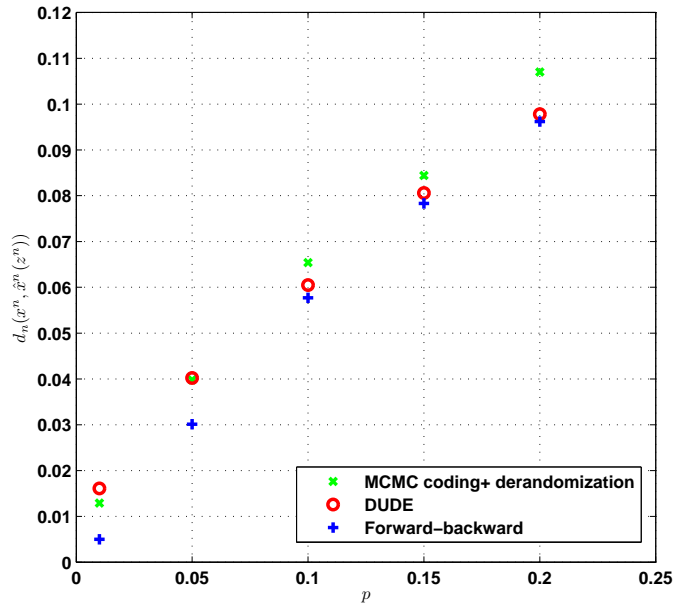}
\caption{Comparing the denoiser based on MCMC coding plus de-randomization with {\sf \footnotesize{DUDE}} and optimal non-universal Bayesian denoiser which is implemented via forward-backward dynamic programming. The source is a BSMS($p$), and the channel is assumed to be a DMC with transition probability $\delta=0.1$. The {\sf \footnotesize{DUDE}} parameters are: $k_{\textmd{letf}}=k_{\textmd{right}}=4$, and the MCMC coder uses $\a=0.9$, $\beta_t=0.5\log t$, $r=10n$, $n=1e4$, $k=7$. The de-randomization window length is $2\times 4 +1=9$.}
\label{fig: 1d denoising}
\end{center}
\end{figure}

As a first example consider a BSMS with transition probability $p$. Fig.~\ref{fig: 1d denoising} compares the performance of {\sf \footnotesize{DUDE}} with the described algorithm. The slope $\alpha$ is chosen such that the expected distortion between the noisy image and its quantized version using Alg.~\ref{alg: mcmc_lossy_coder} is close to the channel probability of error which is $\delta=0.1$ in our case. Here we picked $\alpha=0.9$ for all values of $p$ and did not tune it specifically each time. Though, it can be observed that, even without optimizing the MCMC parameters, the two algorithms have similar performances, and in fact for small values of $p$ the new algorithm outperforms {\sf \footnotesize DUDE}.

\begin{figure}
\begin{center}
\includegraphics[width=25mm]{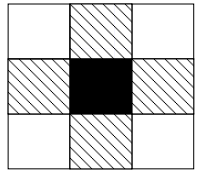}
\caption{The $4^{\rm th}$ order context used by {\sf\footnotesize DUDE}  in 2D image denoising example} \label{fig: grid2}
\end{center}
\end{figure}

\begin{figure}[h]
\begin{center}
\includegraphics[width=25mm]{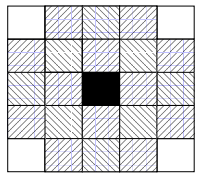}
\caption{The de-randomization block used in MCMC-based denoising of a 2-D image example}\label{fig: grid3}
\end{center}
\end{figure}

\begin{figure}
  \begin{center}
  \includegraphics[width=9cm]{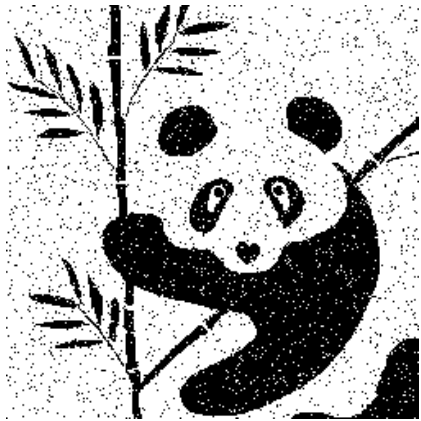} 
   \end{center}
   \caption{Noisy image corrupted by a BSC(0.04)}    \label{fig: noisy image}
\end{figure}

\begin{figure}
  \begin{center}
   \subfigure[{\sf \footnotesize{DUDE}} reconstruction image with $d_N(x^N,\hat{x}^N)=0.0081$: $k_{\textmd{letf}}=k_{\textmd{right}}=4$.]{
    \label{fig: image_hat_DUDE} \includegraphics[width=9cm]{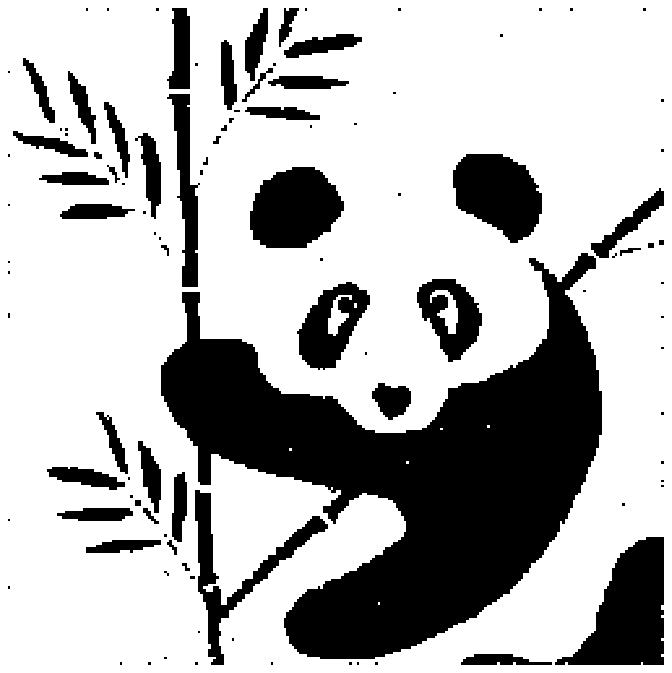} }
    \subfigure[MCMC coder + de-randomization reconstruction image with $d_N(x^N,\hat{x}^N)=0.0128$: $\a=2$, $\beta_t=5\log t$, $r=10N$,] {
    \label{fig: image_hat_MCMC} \includegraphics[width=9cm]{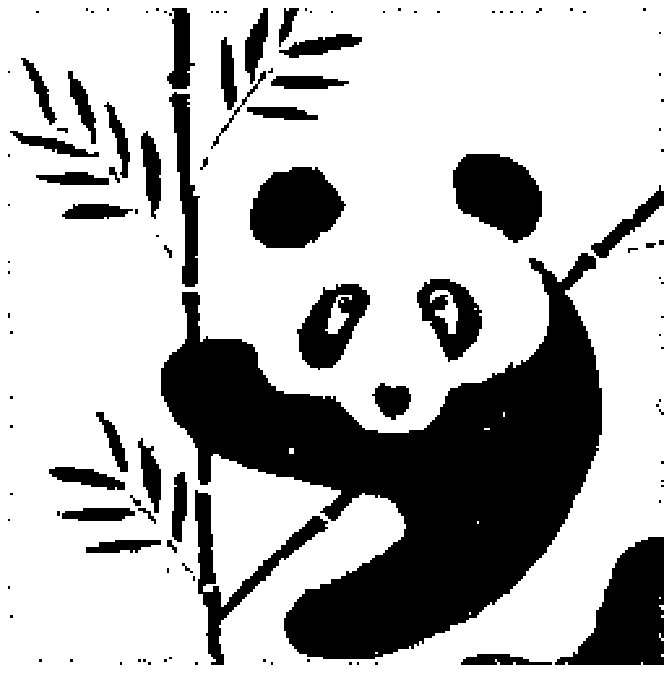} }
    \caption{Comparing the performance of MCMC-based denoiser  with the performance of DUDE}
   \end{center}
\end{figure}

In another example, let us consider denoising the binary image shown in Fig.~\ref{fig: a}. Fig.~\ref{fig: noisy image} shows its noisy version which is generated by passing the original image through a DMC with error probability of $0.04$. Fig.~\ref{fig: image_hat_DUDE} shows the reconstructed image generated by {\sf\footnotesize DUDE} and \ref{fig: image_hat_MCMC} depicts the reconstructed image using the described algorithm. In this experiment the {\sf\footnotesize DUDE} context structure is set as Fig.~\ref{fig: grid2}.
The 2D MCMC coder employs the same context as the one used in the example of Section \ref{sec: sub1} which is shown in Fig.~\ref{fig: grid1}, and the derandomization block is chosen as Fig.~\ref{fig: grid3}.

{\it Discussion:} The new proposed approach which is based on MCMC coding plus de-randomization is an alternative not only to the {\sf\footnotesize  DUDE}, but also to MCMC-based denoising schemes that have been based on and inspired by the Geman brothers' work \cite{Gibbs_sampler}. While algorithmically, this approach has much of the flavor of previous MCMC-based denoising approaches, ours has the merit of leading to a universal scheme, whereas the previous MCMC-based schemes guarantee, at best, convergence to something which is good according to the posterior distribution of the original given the noisy data, but as would be induced by the rather arbitrary  prior model placed on the data. It is clear that here no assumption about the distribution/model of the original data is made.

\section{conclusions and future work}\label{conclusion}

In this paper, a new implementable universal lossy source coding  algorithm based on simulated annealing Gibbs sampling is proposed, and it is shown that it is capable of getting arbitrarily  closely to the rate-distortion curve of any stationary ergodic source. For coding a source sequence $x^n$, the algorithm starts from some initial reconstruction block, and updates one of its coordinates at each iteration. The algorithm can be viewed as a process of systematically introducing `noise' into the original source block, but in a biased direction that results in a decrease of its description complexity. We further developed the application of this new method to universal denoising.

In practice, the proposed algorithms \ref{alg: mcmc_lossy_coder} and \ref{alg: find_f_mcmc_SB}, in their present form, are only applicable to the cases where the size of the reconstruction alphabet, $|\hat{\Xc}|$, is small. The reason is twofold: first, for larger alphabet sizes the contexts will be too sparse to give a true estimate of the empirical entropy of the reconstruction block, even for small values of $k$. Second, the size of the count matrix $\mathbf{m}$ grows exponentially with $|\hat{\Xc}|$ which makes storing it for large values of $|\hat{\Xc}|$ impractical. Despite this fact, there are practical applications where this constraint is satisfied. An example is lossy compression of binary images, like the one presented in Section \ref{sec: simulations}.  Another application for lossy compression of binary data is shown in \cite{defect_list_compression} where one needs to compress a stream of $0$ and $1$ bits with some distortion.

The convergence rate of the new algorithms and the effect of different parameters on it is a topic for further study. As an example, one might wonder how the convergence rate of the algorithm is affected by choosing an initial point other than the source output block itself. Although our theoretical results on universal asymptotic optimality remain intact for any initial starting point, in practice the choice of the starting point might significantly impact the number of iterations required.

Finally, note that in the non-universal setup, where the optimal achievable rate-distortion tradeoff is known in advance, this extra information can be used as a stopping criterion for the algorithm. For example, we can set it to stop after reaching optimum performance to within some fixed distance.




\renewcommand{\theequation}{A-\arabic{equation}}
\setcounter{equation}{0}  

\section*{APPENDIX A: Proof of Theorem \ref{th: the MCMC based scheme}}  \label{app2}

Our proof follows the results presented in \cite{book:MarkovChains}. Throughout this section, $\Ac_n={\hat{\Xc}}^N$ denotes the state space of our Markov chain (MC), $\textbf{P}$ defines a stochastic transition matrix from $\Ac_n$ to itself, and $\boldsymbol\pi$ defines a distribution on $\Ac_n$ satisfying $\boldsymbol\pi\mathbf{P}=\boldsymbol\pi$. Let $N\triangleq{|\mathcal{X}|^n}$ denote the size of the state space, and, for $i\in\{1,\ldots,N\}$,  let $\pb_i$ represent the $i^{\rm th}$ row of $\textbf{P}$.

\begin{definition}[Ergodic coefficient]
Dobrushin's ergodic coefficient of $\textbf{P}$ , $\delta(\textbf{P})$, is defined to be
\begin{equation}
\delta(\textbf{P}) = \max\limits_{1\leq i,j \leq N} \|\mathbf{p}_i-\mathbf{p}_j\|_{\rm TV}.
\end{equation}
From the definition, $0\leq\delta(\textbf{P})\leq 1$. Moreover, since $\|\mathbf{p}_i-\mathbf{p}_j\|_{\rm TV} = 1- \sum\limits_{k=1}^N \min(p_{ik},p_{jk})$, the ergodic coefficient can alternatively be defined as
\begin{equation}\label{eq: alternative definition of Erg. coeff.}
\delta(\textbf{P}) = 1-\min\limits_{1\leq i,j \leq N} \sum_{k=1}^N\min(p_{ik},p_{jk}).
\end{equation}

\end{definition}
The following theorem states the connection between the ergodic coefficient of a stochastic matrix and its convergence rate to the stationary distribution.
\begin{theorem}[Convergence rate in terms of Dobrushin's coefficient] Let $\boldsymbol\mu$ and $\boldsymbol\nu$ be two probability distributions on $\Ac_n$. Then
\begin{align}\label{eq: connection of ergodic coeff. and convergence rate}
\|\boldsymbol\mu\textbf{P}^t-\boldsymbol\nu\textbf{P}^t\|_{1} \leq \|\boldsymbol\mu-\boldsymbol\nu\|_{1}\delta(\textbf{P})^t
\end{align}
\end{theorem}
\begin{corollary}
By substituting $\boldsymbol\nu=\boldsymbol\pi$ in (\ref{eq: connection of ergodic coeff. and convergence rate}), we get
$\|\boldsymbol\mu\textbf{P}^t-\boldsymbol\pi\|_{1} \leq \|\boldsymbol\mu-\boldsymbol\pi\|_{1}\delta(\textbf{P})^t$.
\end{corollary}

Thus far, we talked about homogenous MCs with stationary transition matrix. However, in simulated annealing we deal with a nonhomogeneous MC. The transition probabilities of a nonhomogeneous MC depend on time and vary as time proceeds. Let $\mathbf{P}^{(t)}$ denote the transition Matrix of the MC at time $t$, and for $0\leq n_1<n_2\in \mathds{N}$, define
$\mathbf{P}^{(n_1,n_2)} \triangleq \prod_{t=n_1}^{n_2-1}\mathbf{P}^{(t)}$.
By this definition, if  at time $n_1$ the distribution of the MC on the state space $\Ac_n$ is $\boldsymbol\mu_{n_1}$, at time $n_2$, the distribution evolves to $\boldsymbol\mu_{n_2}=\boldsymbol\mu_{n_1}\mathbf{P}^{(n_1,n_2)}$. The following two definitions characterize the steady state behavior of a nonhomogeneous MC.

\begin{definition}[Weak ergodicity]
A  nonhomogeneous MC is called weakly ergodic if for any distributions $\boldsymbol\mu$ and $\boldsymbol\nu$ over $\Ac_n$, and any $n_1\in\mathds{N}$,
\begin{align}\label{eq: def weak ergodicity}
\limsup\limits_{n_2\to\infty} \|\boldsymbol\mu\mathbf{P}^{(n_1,n_2)}-\boldsymbol\nu\mathbf{P}^{(n_1,n_2)}\|_{1} = 0.
\end{align}
\end{definition}

\begin{definition}[Strong ergodicity]
A  nonhomogeneous Markov chain is called strongly ergodic if there exists a distribution over the state space $\Ac_n$ such that for any distributions $\mu$ and $n_1\in\mathds{N}$,
\begin{align}
\limsup\limits_{n_2\rightarrow \infty} \|\boldsymbol\mu\mathbf{P}^{(n_1,n_2)}-\boldsymbol\pi\|_{1} = 0.
\end{align}
\end{definition}

\begin{theorem}[Block criterion for weak ergodicity] \label{thm: Block criterion for weak ergodicity}
A MC is weakly ergodic iff there exists a sequence of integers $0\leq n_1<n_2<\ldots$, such that
\begin{align}
\sum\limits_{i=1}^{\infty}(1-\delta(\mathbf{P}^{(n_i,n_{i+1})}))=\infty.
\end{align}
\end{theorem}

\begin{theorem}[Sufficient condition for strong ergodicity] \label{thm: Sufficient condition for strong ergodicity}
Let the MC be weakly ergodic. Assume that there exists a sequence of probability distributions, $\{\boldsymbol\pi^{(i)}\}_{i=1}^{\infty}$, on $\Ac_n$ such that $\boldsymbol\pi^{(i)}\mathbf{P}^{(i)}=\boldsymbol\pi^{(i)}$.  Then the MC is strongly ergodic, if
\begin{align}
\sum\limits_{i=1}^{\infty}\|\boldsymbol\pi^{(i)}-\boldsymbol\pi^{(i+1)}\|_{1}<\infty.
\end{align}
\end{theorem}

After stating all the required definitions and theorems from \cite{book:MarkovChains}, finally we get back to our main goal which was to prove that by the mentioned choice of the $\{\beta_t\}$ sequence, Algorithm \ref{alg: mcmc_lossy_coder} converges to the optimal solution asymptotically as block length goes to infinity. Here $\mathbf{P}^{(j)}$, the transition matrix of the MC at the $j^{\rm th}$ iteration, depends on $\beta_j$. Using Theorem \ref{thm: Block criterion for weak ergodicity}, first we prove that the MC is weakly ergodic.

\begin{lemma} \label{lemma: bounding delta}
The ergodic coefficient of $\textbf{P}^{(jn,(j+1)n)}$, for any $j\geq 0$ is upper-bounded as follows
\begin{align}
\delta(\textbf{P}^{(jn,(j+1)n)}) \leq 1-e^{-n(\bar{\b_j}\Delta+\e_n)},
\end{align}
where
\begin{align}\label{eq: def of Delta}
\Delta = \max\limits_{i\in\{1,\ldots,n\}} \delta_i,
\end{align}
for
\[
\delta_i = \max \{|\Ec(u^{i-1}au_{i+1}^n)-\Ec(u^{i-1}bu_{i+1}^n)|;\; u^{i-1} \in \hat{\Xc}^{i-1}, u_{i+1}^n \in \hat{\Xc}^{n-i} , a,b \in \hat{\Xc}  \}.
\]
and
\begin{align}\label{eq: def of e_n}
\e_n = \log n - \frac{\log(n!)}{n}.
\end{align}
\end{lemma}

\begin{proof}
Let $y_1^n$ and $y_2^n$ be two arbitrary sequences in $\hat{\Xc}^n$. Since the Hamming distance between these two sequence is at most $n$, starting from any sequence $y_1^n$, after at most $n$ steps of the Gibbs sampler, it is possible to get to any other sequence $y_2^n$. On the other hand at each step the transition probabilities of jumping from one state to a neighboring state, i.e.,
\begin{align}\label{eq: def on transition prob}
\mathbf{P}^{(t)}(u^{i-1}bu_{i+1}^n|u^{i-1}au_{i+1}^n) = \frac{\exp(-\beta_t\Ec(u^{i-1}au_{i+1}^n))}{n\sum\limits_{b\in\hat{\Xc}}\exp(-\beta_t\Ec(u^{i-1}bu_{i+1}^n))},
\end{align}
can be upper bounded as follows. Dividing both the numerator and denominator of (\ref{eq: def on transition prob}) by $\exp(-\beta_t \Ec_{\min,i}(u^{i-1},u_{i+1}^n))$, where $\Ec_{\min,i}(u^{i-1},u_{i+1}^n)=\min\limits_{b\in\hat{\Xc}}\Ec(u^{i-1}bu_{i+1}^n)$, we get
\begin{align}
\mathbf{P}^{(t)}(u^{i-1}bu_{i+1}^n|u^{i-1}au_{i+1}^n) &= \frac{\exp(-\beta_t(\Ec(u^{i-1}au_{i+1}^n)-\Ec_{\min,i}(u^{i-1},u_{i+1}^n)))}{n\sum\limits_{b\in\hat{\Xc}}\exp(-\beta_t(\Ec(u^{i-1}bu_{i+1}^n)-\Ec_{\min,i}(u^{i-1},u_{i+1}^n)))},\\
&\geq \frac{e^{-\beta_t\Delta}}{n|\hat{\Xc}|}.
\end{align}
Therefore,
\begin{align}
\min\limits_{y_1^n,y_2^n\in\hat{\Xc}^n}\textbf{P}^{(jn,(j+1)n)}(y_1^n,y_2^n) \geq \frac{n!}{n^n}\prod\limits_{t=jn}^{jn+n-1}\frac{e^{-\beta_t\Delta}}{|\hat{\Xc}|} = \frac{e^{-n(\bar{\beta_j}\Delta+\e_n)}}{|\hat{\Xc}|^n},
\end{align}
where $\bar{\beta_j}=\frac{1}{n}\sum\limits_{t=jn}^{jn+n-1}\beta_t$.

Using the alternative definition of the ergodic coefficient given in (\ref{eq: alternative definition of Erg. coeff.}),
\begin{align}
\delta(\textbf{P}^{(jn,(j+1)n)}) &= 1-\min\limits_{y_1^n,y_2^n\in\hat{\Xc}^n} \sum\limits_{z^n\in\hat{\Xc}^n}\min(\textbf{P}^{(jn,(j+1)n)}(y_1^n,z^n),\textbf{P}^{(jn,(j+1)n)}(y_2^n,z^n)) \nonumber\\
&\leq 1-|\hat{\Xc}|^n\frac{1}{|\hat{\Xc}|^n}e^{-n(\bar{\beta_j}\Delta+\e_n)}\\
&= 1-e^{-n(\bar{\beta_j}\Delta+\e_n)}.
\end{align}
\end{proof}

\begin{corollary}
Let $\beta_t=\frac{\log(\lfloor \frac{t}{n} \rfloor+1)}{T_0^{(n)}}$, where $T_0^{(n)}=cn\Delta$, for some $c>1$, and $\Delta$ is defined in (\ref{eq: def of Delta}), in Algorithm \ref{alg: mcmc_lossy_coder}. Then the generated MC is weakly ergodic.
\end{corollary}
\begin{proof}
 For proving weak ergodicity, we use the block criterion stated in Theorem \ref{thm: Block criterion for weak ergodicity}. Let $n_j=jn$, and note that $\bar{\beta_j}=\frac{\log (j+1)}{T_0}$ in this case. Observe that
\begin{align}
\sum\limits_{j=0}^{\infty}(1-\delta(\mathbf{P}^{(n_j,n_{j+1})}))&=\sum\limits_{j=1}^{\infty}(1-\delta(\mathbf{P}^{(jn,(j+1)n)}))\nonumber\\
&\geq \sum_{j=0}^{\infty}e^{-n(\bar{\beta_j}\Delta+\e_n)}\\
&= \sum_{j=0}^{\infty}e^{-n(\Delta\frac{\log(j+1)}{T_0}+\e_n)}\\
&= e^{-n\e_n}\sum_{j=1}^{\infty}\frac{1}{j^{1/c}}=\infty.
\end{align}
This yields the weak ergodicity of the MC defined by the simulated annealing and Gibbs sampler.
\end{proof}

Now we are ready to prove the result stated in Theorem \ref{th: the MCMC based scheme}. Using Theorem \ref{thm: Sufficient condition for strong ergodicity},  we prove that the MC is in fact strongly ergodic and the eventual steady state distribution of the MC as the number of iterations converge to infinity is  a uniform distribution over the sequences that minimize the energy function.

At each time $t$, the distribution defined as $ \boldsymbol\pi^{(t)}(y^n)=e^{-\beta_t\Ec(y^n)}/Z_{\beta_t}$ satisfies $\boldsymbol\pi^{(t)}\mathbf{P}^{(t)}=\boldsymbol\pi^{(t)}$. Therefore, if we prove that
\begin{align}\label{eq: required condition}
\sum_{t=1}^{\infty}\|\boldsymbol\pi^{(t)}-\boldsymbol\pi^{(t+1)}\|_{1}<\infty,
\end{align}
by Theorem \ref{thm: Sufficient condition for strong ergodicity}, the MC is also strongly ergodic. But it is easy to show that $\boldsymbol\pi^{(t)}$ converges to a uniforms distribution over the set of sequences that minimize the energy function, i.e.,
\begin{align}
\lim_{t\rightarrow\infty}\boldsymbol\pi^{(t)}(y^n)=\left\{\begin{array}{c}
                                                0 ; \quad y^n \notin \Hc,\\
                                                \frac{1}{|\Hc|} ; \quad y^n \in \Hc,
                                              \end{array}
\right.
\end{align}
where $\Hc\triangleq\{y^n:\Ec( y^n )=\min\limits_{z^n\in\hat{\Xc}^n}\Ec(z^n)\}$.

Hence, if we let $\hat{X}^n_t$ denote the output of Algorithm \ref{alg: mcmc_lossy_coder} after $t$ iterations, then
\begin{align}
\lim\limits_{t\rightarrow\infty}\Ec(\hat{X}^n_t)=\min_{y^n\in \hat{\Xc}^n} \Ec(y^n),
\end{align}
which combined with Theorem \ref{th: the exhaustive search scheme} yields the desired result.

In order to prove (\ref{eq: required condition}), we prove that $\boldsymbol\pi^{(t)}(y^n)$ is increasing on $\Hc$, and eventually decreasing on $\Hc^c$, hence there exists $t_0$ such that for any $t_1>t_0$,
\begin{align}
\sum_{t=t_0}^{t_1}\|\boldsymbol\pi^{(t_1)}-\boldsymbol\pi^{(t+1)}\|_{1}&=
\frac{1}{2}\sum\limits_{y^n\in \Hc}\sum_{t=t_0}^{t_1}(\boldsymbol\pi^{(t+1)}(y^n)-\boldsymbol\pi^{(t)}(y^n))
+\frac{1}{2}\sum\limits_{y^n\in\hat{\Xc}^n\backslash \Hc} \sum_{t=t_0}^{t_1}(\boldsymbol\pi^{(t)}(y^n)-\boldsymbol\pi^{(t+1)}(y^n)),\nonumber\\
&=\frac{1}{2}\sum\limits_{y^n\in \Hc}(\boldsymbol\pi^{(t_1+1)}(y^n)-\boldsymbol\pi^{(t_0)}(y^n))
+\frac{1}{2}\sum\limits_{y^n\in\hat{\Xc}^n\backslash \Hc} (\boldsymbol\pi^{(t_0)}(y^n)-\boldsymbol\pi^{(t_1+1)}(y^n)),\nonumber\\
&<\frac{1}{2}(1). \label{eq:finite}
\end{align}
Since the right hand side of \eqref{eq:finite} of does not depend on $t_1$, $ \sum_{t=0}^{\infty}\|\boldsymbol\pi^{(t)}-\boldsymbol\pi^{(t+1)}\|_{1}<\infty.$ Finally, in order to prove that  $\boldsymbol\pi^{(t)}(y^n)$ is increasing for $y^n\in \Hc$, note that
\begin{align}
\boldsymbol\pi^{(t)}(y^n) &= \frac{e^{-\beta_t\Ec(y^n)}}{\sum\limits_{z^n\in\hat{\Xc}^n}e^{-\beta_t\Ec(z^n)}}\nonumber\\
               &= \frac{1}{\sum\limits_{z^n\in\hat{\Xc}^n}e^{-\beta_t(\Ec(z^n)-\Ec(y^n))}}.\label{eq: increasing}
\end{align}
Since for $y^n\in \Hc$ and any $z^n\in\hat{\Xc}^n$, $\Ec(z^n)-\Ec(y^n)\geq0$, if $t_1<t_2$,
\[
\sum\limits_{z^n\in\hat{\Xc}^n}e^{-\beta_{t_1}(\Ec(z^n)-\Ec(y^n))}>\sum\limits_{z^n\in\hat{\Xc}^n}e^{-\beta_{t_2}(\Ec(z^n)-\Ec(y^n))},
\]
and hence $\boldsymbol\pi^{(t_1)}(y^n)<\boldsymbol\pi^{(t_2)}(y^n)$. On the other hand,  if $y^n\notin \Hc$, then
\begin{align}
\boldsymbol\pi^{(t)}(y^n) &= \frac{e^{-\beta_t\Ec(y^n)}}{\sum\limits_{z^n\in\hat{\Xc}^n}e^{-\beta_t\Ec(z^n)}}\nonumber\\
               &= \frac{1}{\sum\limits_{z^n:\Ec(z^n)\geq\Ec(y^n)}e^{-\beta_t(\Ec(z^n)-\Ec(y^n))}+\sum\limits_{z^n:\Ec(z^n)<\Ec(y^n)}e^{\beta_t(\Ec(y^n)-\Ec(z^n))}}. \label{eq: decreasing}
\end{align}
For large $\beta$ the denominator of (\ref{eq: decreasing}) is dominated by the second term which is increasing in $\beta_t$ and therefore $\boldsymbol\pi^{(t)}(y^n)$ will be decreasing in $t$. This concludes the proof.

\renewcommand{\theequation}{B-\arabic{equation}}
\setcounter{equation}{0}  

\section*{APPENDIX B: Proof of Theorem \ref{th: the MCMC based scheme for finding SB code}}  \label{app3}

First we need to prove that a result similar to Theorem \ref{th: the exhaustive search scheme} holds for SB codes. I.e., we need to prove that for given sequences   $\{k_f^{(n)}\}_n$ and $\{k_n\}_n$  such that $\lim\limits_{n\rightarrow \infty}k_f^{(n)}=\infty$ and  $k_n=o(\log n)$, finding a sequence of SB codes according to
\begin{align}
\hat{f}^{K_f^{(n)}} = \argmin\limits_{f^{K_f^{(n)}}}\Ec(f^{K_f^{(n)}}),
\end{align}
where $\Ec(f^{K_f^{(n)}})$ is defined in (\ref{eq: energy function to SB code}) and $K_f^{(n)}=2^{2k_f^{(n)}+1}$, results in a sequence of asymptotically optimal codes for any stationary ergodic source $\mathbf{X}$ at slope $\a$. In other words,
\begin{align}\label{eq: thm 3 claim}
\lim\limits_{n\rightarrow\infty} \left[\frac{1}{n}\ell_{{\sf\footnotesize  LZ}}(\hat{X}^n)+\a d_n(\hat{X}^n,X^n)\right] = \min\limits_{D\geq0}\left[R(D,\mathbf{X})+\a D\right], \quad\textmd{a.s.}
\end{align}
where $\hat{X}^n = \hat{X}^n[X^n,\hat{f}^{K_f^{(n)}}]$. After proving this, the rest of the proof follows from the proof of Theorem \ref{th: the MCMC based scheme} by just redefining $\delta_i$ as
\[
\delta_i = \max \left\{\left|\Ec(f^{i-1}af_{i+1}^{K_f^{(n)}})-\Ec(f^{i-1}bf_{i+1}^{K_f^{(n)}})\right|;\; f^{i-1} \in \hat{\Xc}^{i-1}, f_{i+1}^{K_f^{(n)}} \in \hat{\Xc}^{K_f^{(n)}-i} , a,b \in \hat{\Xc} \right\}.
\]

For establishing the equality stated in (\ref{eq: thm 3 claim}), similar to the proof of Theorem \ref{th: the exhaustive search scheme}, we prove consistent lower and upper bounds which in the limit yield the desired result. The lower bound,
\begin{align}
\liminf\limits_{n\rightarrow\infty}\E\left[ \frac{1}{n} \ell_{{\sf\footnotesize  LZ}} (\hat{X}^n)+\a d(X^n,\hat{X}^n)\right]\geq \min\limits_{D\geq0}\left[R(D,\mathbf{X})+\a D\right],\label{eq: SB codes: lower bound on cost function}
\end{align}
follows from part (1) of Theorem 5 in \cite{YangZ:97}. For proving the upper bound,  we split the cost into two terms, as done in the equation (\ref{eq: thm 1: split cost}). The convergence to zero of the first term again follows from a similar argument. The only difference is in upper bounding the second term.

Since, asymptotically, for any stationary ergodic process $\mathbf{X}$, SB codes have the same rate-distortion performance as block codes, for a point $(R(D,\mathbf{X}),D)$ on the rate-distortion curve of the source, and any $\epsilon>0$, there exits a SB code $f^{2\kappa_f^{\epsilon}+1}$ of  some order $\kappa_f^{\epsilon}$ such that coding the process $\mathbf{X}$ by this SB code results in a process $\tilde{\mathbf{X}}$ which satisfies
\begin{enumerate}
\item $\bar{H}(\tilde{\mathbf{X}})\leq R(D,\mathbf{X}),$
\item $\E d(X_0,\tilde{X}_0)\leq D+\epsilon$.
\end{enumerate}
On the other hand, for a fixed $n$, $\Ec(f^{K_f})$ is monotonically decreasing in $K_f$. Therefore, for any process $\mathbf{X}$ and any $\delta>0$, there exists $n_{\delta}$ such that for $n>n_{\delta}$ and $k_f^{(n)}\geq\kappa_f^{\epsilon}$
\begin{align}\label{eq: SB upper}
\limsup\limits_{n\rightarrow\infty}\left[H_{k_n}(\hat{X}^n)+\a d_n(X^n,\hat{X}^n)\right]\leq  R(D,\mathbf{X})+\a(D+\epsilon)+\delta,\quad\textmd{w.p. 1}.
\end{align}

Combining  (\ref{eq: SB codes: lower bound on cost function}) and (\ref{eq: SB upper}), plus the arbitrariness of $\epsilon$, $\delta$ and $D$ yield the desired result.

\bibliographystyle{unsrt}

\bibliography{myrefs}

\begin{thebibliography}{10}

\bibitem{cover}
T.~Cover and J.~Thomas.
\newblock {\em Elements of Information Theory}.
\newblock Wiley, New York, 2nd edition, 2006.

\bibitem{Shannon48}
C.~E. Shannon.
\newblock A mathematical theory of communication.
\newblock {\em Bell Syst. Tech. J.}, 27:379--423 and 623--656, 1948.

\bibitem{Gallager}
R.G. Gallager.
\newblock {\em Information Theory and Reliable Communication}.
\newblock NY: John Wiley, 1968.

\bibitem{book:Berger}
T.~Berger.
\newblock {\em Rate-distortion theory: A mathematical basis for data
  compression}.
\newblock NJ: Prentice-Hall, 1971.

\bibitem{LZ}
J.~Ziv and A.~Lempel.
\newblock Compression of individual sequences via variable-rate coding.
\newblock {\em Information Theory, IEEE Transactions on}, 24(5):530--536, Sep
  1978.

\bibitem{arith_coding}
I.~H. Witten, R.~M. Neal, , and J.~G. Cleary.
\newblock Arithmetic coding for data compression.
\newblock {\em Commun. Assoc. Comp. Mach.}, 30(6):520--540, 1987.

\bibitem{ISIT_JW}
S.~Jalali and T.~Weissman.
\newblock New bounds on the rate-distortion function of a binary markov source.
\newblock In {\em Information Theory, 2007. ISIT 2007. IEEE International
  Symposium on}, pages 571--575, June 2007.

\bibitem{GrayN:75}
R.~Gray, D.~Neuhoff, and J.~Omura.
\newblock Process definitions of distortion-rate functions and source coding
  theorems.
\newblock {\em Information Theory, IEEE Transactions on}, 21(5):524--532, Sep
  1975.

\bibitem{simulated_annealing_1}
S.~Kirkpatrick, C.~D. Gelatt, Jr., and M.~P. Vecchi.
\newblock Optimization by simulated annealing.
\newblock {\em Science}, 220:671--680, 1983.

\bibitem{simulated_annealing_2}
V.~Cerny.
\newblock Thermodynamical approach to the traveling salesman problem: An
  efficient simulation algorithm.
\newblock {\em Journal of Optimization Theory and Applications}, 45(1):41--51,
  Jan 1985.

\bibitem{Gibbs_sampler}
S.~Geman and D.~Geman.
\newblock Stochastic relaxation, gibbs distributions and the bayesian
  restoration of images.
\newblock {\em IEEE Transactions on Pattern Analysis and Machine Intelligence},
  6:721--741, Nov 1984.

\bibitem{deterministic_annealing}
Kenneth Rose.
\newblock Deterministic annealing for clustering, compression, classification,
  regression, and related optimization problems.
\newblock {\em Proceedings of the IEEE}, 86(11):2210--2239, Nov 1998.

\bibitem{SA_codebook_design}
J.~Vaisey and A.~Gersho.
\newblock Simulated annealing and codebook design.
\newblock In {\em Acoustics, Speech, and Signal Processing, 1988. ICASSP-88.,
  1988 International Conference on}, pages 1176--1179 vol.2, Apr 1988.

\bibitem{GLA}
Y.~Linde, A.~Buzo, and R.~Gray.
\newblock An algorithm for vector quantizer design.
\newblock {\em Communications, IEEE Transactions on}, 28(1):84--95, Jan 1980.

\bibitem{Sakrison:70}
D.~J. Sakrison.
\newblock The rate of a class of random processes.
\newblock {\em Information Theory, IEEE Transactions on}, 16:10--16, Jan. 1970.

\bibitem{Ziv:72}
J.~Ziv.
\newblock Coding of sources with unknown statisticsÑpart ii: Distortion
  relative to a fidelity criterion.
\newblock {\em Information Theory, IEEE Transactions on}, 18:389--394, May
  1972.

\bibitem{NeuhoffG:75}
D.~L. Neuhoff, R.~M. Gray, and L.D. Davisson.
\newblock Fixed rate universal block source coding with a fidelity criterion.
\newblock {\em Information Theory, IEEE Transactions on}, 21:511--523, May
  1972.

\bibitem{NeuhoffS:78}
D.~L. Neuhoff and P.~L. Shields.
\newblock Fixed-rate universal codes for markov sources.
\newblock {\em Information Theory, IEEE Transactions on}, 24:360--367, May
  1978.

\bibitem{Ziv:80}
J.~Ziv.
\newblock Distortion-rate theory for individual sequences.
\newblock {\em Information Theory, IEEE Transactions on}, 24:137--143, Jan.
  1980.

\bibitem{GarciaN:82}
R.~Garcia-Munoz and D.~L. Neuhoff.
\newblock Strong universal source coding subject to a rate-distortion
  constraint.
\newblock {\em Information Theory, IEEE Transactions on}, 28:285Ð295, Mar.
  1982.

\bibitem{CheungS:90}
K.~Cheung and V.~K. Wei.
\newblock A locally adaptive source coding scheme.
\newblock {\em Proc. Bilkent Conf on New Trends in Communication, Control, and
  Signal Processing}, pages 1473--1482, 1990.

\bibitem{BentleyS:86}
J.L. Bentley, D.D. Sleator, R.E. Tarjan, and V.K. Wei.
\newblock A locally adaptive data compression algorithm.
\newblock {\em Communications of the ACM}, 29(4):320--330, Apr 1986.

\bibitem{MoritaK:89}
H.~Morita and K.~Kobayashi.
\newblock An extension of lzw coding algorithm to source coding subject to a
  fidelity criterion.
\newblock In {\em In Proc. 4th Joint Swedish-Soviet Int. Workshop on
  Information Theory}, page 105–109, Gotland, Sweden, 1989.

\bibitem{SteinbergG:93}
Y.~Steinberg and M.~Gutman.
\newblock An algorithm for source coding subject to a fidelity criterion based
  on string matching.
\newblock {\em Information Theory, IEEE Transactions on}, 39:877Ð886, Mar.
  1993.

\bibitem{YangK:98}
En~hui Yang and J.C. Kieffer.
\newblock On the performance of data compression algorithms based upon string
  matching.
\newblock {\em Information Theory, IEEE Transactions on}, 44(1):47 --65, jan
  1998.

\bibitem{LuczakS:97}
T.~Luczak and T.~Szpankowski.
\newblock A suboptimal lossy data compression based on approximate pattern
  matching.
\newblock {\em Information Theory, IEEE Transactions on}, 43:1439Ð1451, Sep.
  1997.

\bibitem{ZhangW:96}
Zhen Zhang and V.K. Wei.
\newblock An on-line universal lossy data compression algorithm via continuous
  codebook refinement. i. basic results.
\newblock {\em Information Theory, IEEE Transactions on}, 42(3):803 --821, may
  1996.

\bibitem{kontoyiannis_1}
I.~Kontoyiannis.
\newblock An implementable lossy version of the lempel ziv algorithm-part i:
  optimality for memoryless sources.
\newblock {\em Information Theory, IEEE Transactions on}, 45(7):2293--2305, Nov
  1999.

\bibitem{OrnsteinS:90}
D.~S. Ornstein and P.~C. Shields.
\newblock Universal almost sure data compression.
\newblock {\em The Annals of Probability}, pages 441--452, Mar. 1990.

\bibitem{YangK:96}
En~hui Yang and J.C. Kieffer.
\newblock Simple universal lossy data compression schemes derived from the
  lempel-ziv algorithm.
\newblock {\em Information Theory, IEEE Transactions on}, 42(1):239--245, Jan
  1996.

\bibitem{NeuhoffS:98}
D.~L. Neuhoff and P.~C. Shields.
\newblock Simplistic universal coding.
\newblock {\em Information Theory, IEEE Transactions on}, 44:778--781, Mar.
  1998.

\bibitem{YangZ:97}
En~hui Yang, Z.~Zhang, and T.~Berger.
\newblock Fixed-slope universal lossy data compression.
\newblock {\em Information Theory, IEEE Transactions on}, 43(5):1465--1476, Sep
  1997.

\bibitem{non-universal_R_D_coder_ref1}
M.J. Wainwright and E.~Maneva.
\newblock Lossy source encoding via message-passing and decimation over
  generalized codewords of ldgm codes.
\newblock In {\em Information Theory, 2005. ISIT 2005. Proceedings.
  International Symposium on}, pages 1493--1497, Sept. 2005.

\bibitem{non-universal_R_D_coder_ref2}
A.~Gupta and S.~Verdu.
\newblock Nonlinear sparse-graph codes for lossy compression of discrete
  nonredundant sources.
\newblock In {\em Information Theory Workshop, 2007. ITW '07. IEEE}, pages
  541--546, Sept. 2007.

\bibitem{R_D_coder_ref3}
J.~Rissanen and I.~Tabus.
\newblock Rate-distortion without random codebooks.
\newblock In {\em Workshop on Information Theory and Applications (ITA)}, Sep
  2006.

\bibitem{R_D_coder_ref4}
A.~Gupta, S.~Verd\'{u}, and T.~Weissman.
\newblock Linear-time near-optimal lossy compression.
\newblock In {\em Information Theory, 2008. ISIT 2008. Proceedings.
  International Symposium on}, 2008.

\bibitem{Ziv_inequality}
E.~Plotnik, M.J. Weinberger, and J.~Ziv.
\newblock Upper bounds on the probability of sequences emitted by finite-state
  sources and on the redundancy of the lempel-ziv algorithm.
\newblock {\em Information Theory, IEEE Transactions on}, 38(1):66--72, Jan
  1992.

\bibitem{Gray_Neuhoff_Ornstein75}
Robert~M. Gray, David~L. Neuhoff, and Donald~S. Ornstein.
\newblock Nonblock source coding with a fidelity criterion.
\newblock {\em The Annals of Probability}, 3(3):478--491, Jun 1975.

\bibitem{Marton_SB}
K.~Markon.
\newblock On the rate distortion function of stationary sources.
\newblock {\em Probl. Contr. Inform. Theory}, 4:289--297, 1975.

\bibitem{Gray_SB}
R.~M. Gray.
\newblock Block, sliding-block, and trellis codes.
\newblock In {\em Janos Bolyai Colloquiem on Info. Theory}, Keszthely, Hungary,
  August 1975.

\bibitem{book:Mallat}
S.~Mallat.
\newblock {\em A Wavelet Tour of Signal Processing}.
\newblock Academic Press, Boston, 1997.

\bibitem{Gray_markov_source}
R.~Gray.
\newblock Rate distortion functions for finite-state finite-alphabet markov
  sources.
\newblock {\em Information Theory, IEEE Transactions on}, 17(2):127--134, Mar
  1971.

\bibitem{Natarajan}
B.~Natarajan, K.~Konstantinides, and C.~Herley.
\newblock Occam filters for stochastic sources with application to digital
  images.
\newblock {\em Signal Processing, IEEE Transactions on}, 46(5):1434--1438, May
  1998.

\bibitem{kolmogrov_sampler}
D.~Donoho.
\newblock The kolmogorov sampler, Jan 2002.

\bibitem{Tsachy-Erik-IT-05}
T.~Weissman and E.~Ordentlich.
\newblock The empirical distribution of rate-constrained source codes.
\newblock {\em Information Theory, IEEE Transactions on}, 51(11):3718--3733,
  Nov 2005.

\bibitem{RV:slope}
K.~Ramchandran and M.~Vetterli.
\newblock Best wavelet packet bases in a rate-distortion sense.
\newblock {\em Image Processing, IEEE Transactions on}, 2(2):160--175, Apr
  1993.

\bibitem{Erik_seroussi_verdu_Weinberger_tsachy}
E.~Ordentlich, G.~Seroussi, S.~Verdu, M.~Weinberger, and T.~Weissman.
\newblock A discrete universal denoiser and its application to binary images.
\newblock In {\em Image Processing, 2003. ICIP 2003. Proceedings. 2003
  International Conference on}, volume~1, pages I--117--20 vol.1, Sep 2003.

\bibitem{DUDE}
T.~Weissman, Erik Ordentlich, G.~Seroussi, S.~Verd\'u, and M.~Weinberger.
\newblock Universal discrete denoising: Known channel.
\newblock {\em IEEE Trans. Inform. Theory}, 51(1):5--28, 2005.

\bibitem{defect_list_compression}
G.~Motta, E.~Ordentlich, and M.J. Weinberger.
\newblock Defect list compression.
\newblock In {\em Information Theory, 2008. ISIT 2008. IEEE International
  Symposium on}, pages 1000--1004, July 2008.

\bibitem{book:MarkovChains}
P.~Bremaud.
\newblock {\em Markov chains, Gibbs fields, Monte Carlo simulation, and
  queues}.
\newblock Springer, New York, 1991.

\end{thebibliography}

\end{document}